\documentclass{jpp}
\usepackage{graphicx}

\usepackage[utf8]{inputenc}
\usepackage[T1]{fontenc}
\usepackage{amsmath}
\usepackage{amssymb}
\usepackage{subcaption}
\usepackage{enumitem}
\usepackage{setspace}

\shorttitle{On conservation laws in the drift-reduced Braginskii model}
\shortauthor{S. García Herreros et al.}

\title{On energy conservation laws in the drift-reduced Braginskii model}

\author{S. García Herreros\aff{1}
  \corresp{\email{sergio.garciaherreros@epfl.ch}},
  B. De Lucca\aff{1}, D. Mancini\aff{1}, Z. Tecchiolli\aff{1}, 
  \\M. Bassanini\aff{1,}\aff{2}, P. Ricci \aff{1}, L. Stenger\aff{1},  
 \and C. Theiler\aff{1}}

\affiliation{\aff{1}École Polytechnique Fédérale de Lausanne (EPFL), Swiss Plasma Center (SPC), CH-1015 Lausanne, Switzerland
\aff{2}École Polytechnique Fédérale de Lausanne (EPFL), Institute of Mathematics, CH-1015 Lausanne, Switzerland}

\begin{document}

\maketitle

\begin{abstract}
A revision of the drift-reduced Braginskii model is presented, focusing on its practical implementation in numerical codes, and systematically analyzing the effect of several commonly-used approximations on energy conservation. As a practical example, the effect of the approximations is quantified in the GBS code. By dropping the most important approximations and related energy sinks, the simulation shows increased transport levels and a target heat flux with closer agreement to experimental observations.
\end{abstract}

\section{Introduction} \label{sec:Intro} 

The dynamics of the plasma in the boundary of fusion devices, the region including the edge and scrape-off layer (SOL), is the result of complex multiphysics and multiscale phenomena, with turbulence playing a key role determining particle and energy transport to the plasma-facing components. A number of models have been derived over the years for the study of plasma turbulence in the boundary, ranging from gyrokinetic \citep{Brizard07, Baptiste_2021} to drift-kinetic \citep{Jorge2019} and gyrofluid theories \citep{Madsen13}. However, because of the high-collisionality and its relatively reduced computational complexity, the  boundary region is still most often studied with a fluid approach based on the Braginskii equations \citep{Braginskii65}. Given the large range of time and spatial scales of the plasma dynamics, the drift-reduced limit of the Braginskii model is used in practice for the study of boundary plasmas \citep{Drake84,Zeiler97,Scott03,Simakov03}. A number of codes have been developed to solve the drift-reduced equations, among those we mention GBS \citep{Ricci12,Halpern2015,Giacomin22}, GDB \citep{Ben18}, GRILLIX \citep{Stegmeir19,Zholobenko24}, Hermes-3/BOUT++ \citep{Dudson24,Dudson_2026}, HESEL \citep{hesel}, and SOLEDGE3X \citep{Bufferand21,Quadri24} previously TOKAM3X \citep{Tatali21}.

Thanks to the development and improvement of both the numerical algorithms and the computational power used for plasma boundary simulations in recent years, full-sized tokamak simulations are now possible, as demonstrated in Oliveira \textit{et al.} (2022), and recently also for stellarators \citep{Antonio23,Tecchiolli_2026,Stegmeir}. Qualitative validation of code results shows good agreement with experimental measurements \citep{Mancini24,Lim23,Zholobenko24,Quadri24,Dudson_2026}. These results are encouraging the use of drift-reduced Braginskii codes as predictive tools in the design of fusion power plants, calling for a detailed analysis of the underlying models. 

The formulation of exact conservation laws for the quantities of physical relevance (mass, charge, momentum and energy) is of particular importance when deriving the drift-reduced fluid models. Recent analytical progress to formulate exact conservation laws for the drift-reduced Braginskii system \citep{Brenno25} yields an analytical expression for the polarization velocity by avoiding the perturbative expansion \citep{Zeiler97} or alternative ad-hoc formulations introduced earlier \citep{Halpern23}. Along the same direction, this work provides an analysis of the problems that prevent satisfying exact conservation laws in practice when using numerical codes that solve the drift-reduced Braginskii equations.

The paper is organized as follows. Following the present Introduction, in Section \ref{sec:Braginskii} we revisit the Braginskii equations, in particular to include external sources of density, momentum and energy, and explicitly derive the energy conservation law of the system. In Section \ref{sec:ideal} we focus on the drift reduction, discussing the formulations of the drift-reduced Braginskii model that are usually employed when implemented in numerical codes, explicitly indicating the most common approximations adopted. In Section \ref{sec:reconstructing} we focus specifically on the equations implemented in the GBS code to rederive the energy conservation law. We then estimate the magnitude of the conservation-breaking approximations. In Section \ref{sec:simulationresults} we underline the importance of energy conservation by comparing simulation results with experimental measurements. Finally, we present our conclusions in Section \ref{sec:Conclusion}.

\section{Braginskii model} \label{sec:Braginskii}

We consider a magnetized plasma with two species $s=\{e,i \}$ (electrons and ions). The Boltzmann equation for species $s$ is given by \citep{Braginskii65}

\begin{equation}
\begin{aligned}
    \frac{\partial f_s}{\partial t} + \boldsymbol{v} \cdot \frac{\partial f_s}{\partial \boldsymbol{x}} + q_s \left( \frac{\boldsymbol{E} + \boldsymbol{v} \times \boldsymbol{B}}{m_s} \right) \cdot \frac{\partial f_s}{\partial \boldsymbol{v}} &= C_s+ C_{\textit{sn}} \\+ \frac{e}{| q_s|}S_n\left(\frac{m_s}{2\pi T_s}\right)^{3/2} \exp{\left(-\frac{m_s(\boldsymbol{v}-\boldsymbol{V}_s)^2}{2 T_s}\right)}
    &+ S_p^s\frac{2 m_s}{3 n_s T_s^2} \left( \frac{m_s(v^2-V_s^2)}{2} - \frac{3}{2}T_s \right) f_s
    \label{boltzmann}
\end{aligned}
\end{equation}

\noindent
where $f_s$ is the distribution function of species $s$, $\boldsymbol{x}$ and $\boldsymbol{v}$ are the position and velocity in phase-space, $q_s$ and $m_s$ are the charge and mass of species $s$, $\boldsymbol{E}$ and $\boldsymbol{B}$ are the electric and magnetic fields and $C_s=\sum_{s'}C_{ss'}$ is the collision operator which models $i-i$, $e-e$ and $i-e$ Coulomb collisions. 

With respect to the original Boltzmann equation proposed by Braginskii, we include the interaction with neutrals and external sources. Neutral species \citep{Giacomin22, Zholobenko24, Quadri24} can interact with the plasma species through the $C_{\textit{sn}}$ collision term. Even though we keep track of the effect of plasma-neutral interactions, the discussion of the specific form of this operator and its related terms is outside of the scope of this work. Flux-driven codes also require density and power sources that ultimately drive turbulence \citep{Giacomin22,Zholobenko24,Quadri24,Tatali21,Ben18}. In practice, this is often introduced by means of a density source term $S_n$ and a power source given by the $S_p^s$ term. To introduce these terms we split the velocity as $\boldsymbol{v}_s = \boldsymbol{V}_s + \boldsymbol{w}_s$ with $\boldsymbol{V}_s$ the bulk flow velocity defined as $\boldsymbol{V}_s = \int f_s \boldsymbol{v}\;d\boldsymbol{v}/\int f_s d\boldsymbol{v}$, and $\boldsymbol{w}_s$ the velocity associated with thermal motion. We denote $n_s$ and $T_s$ the density and temperature of species $s$. Note that we do not specify an index $s$ for $S_n$ since the charge neutrality of ionization and recombination processes imposes $\sum_s q_s S_n^s=0$, which for a hydrogenic plasma with $q_i=-q_e=e$, leads to $S_n^i=S_n^e=S_n$. 

The fluid equations are obtained by taking the velocity moments of equation (\ref{boltzmann}). Integrating the first three moments, we retrieve the time evolution equations of particle density $n_s$, momentum density $\boldsymbol{\mathcal{M}}_s=m_s n_s \boldsymbol{V}_s$ and energy density \mbox{$\mathcal{H}_s=m_s n_s V_s^2/2 + 3 n_sT_s/2$}, that is:

\begin{align}
    & \frac{\partial n_s}{\partial t} 
      + \nabla \cdot \left( n_s \boldsymbol{V}_s \right)
      = S_{n} + S_n^{\textit{sn}},
      \label{eqn:continuityonespecies}
    \\[1em]
    & \begin{aligned}
        \frac{\partial \boldsymbol{\mathcal{M}}_s}{\partial t}
        + \nabla \cdot (\boldsymbol{V}_s \boldsymbol{\mathcal{M}}_s +\boldsymbol{P}_s) &= q_s n_s (\boldsymbol{E} + \boldsymbol{V}_s \times \boldsymbol{B})
        + \boldsymbol{R}_s \\
        &\quad + \boldsymbol{\mathcal{M}}_s S_{n}/n_s
        + \boldsymbol{S}_{\boldsymbol{\mathcal{M}}}^{\textit{sn}},
    \end{aligned}
    \label{eqn:momentumonespecies}
    \\[1em]
    & \begin{aligned}
        \frac{\partial \mathcal{H}_s }{\partial t}
        + \nabla \cdot \left( \mathcal{H}_s \boldsymbol{V}_s
        + \boldsymbol{P}_s \cdot \boldsymbol{V}_s
        + \boldsymbol{q}_s \right)
        &= q_s n_s \boldsymbol{V}_s \cdot \boldsymbol{E}
        + Q_s + \boldsymbol{R}_s \cdot \boldsymbol{V}_s \\
        &\quad + \mathcal{H}_s S_n/n_s
        + S_{p}^s + S_p^{\textit{sn}},
    \end{aligned}
    \label{eqn:energyonespecies}
\end{align}

\noindent
where $S_n^{\textit{sn}}=\int C_{\textit{sn}} d\boldsymbol{v}$ is the density source resulting from plasma-neutral interactions (also satisfying $\sum_s q_s S_n^{\textit{sn}}=0$), $\boldsymbol{P}_s = p_s \mathbb{I} + \boldsymbol{\pi}_s$ is the pressure tensor defined in terms of the scalar pressure $p_s=n_sT_s$ and the stress tensor $\boldsymbol{\pi}_s$, $\boldsymbol{R}_s$ and \mbox{$\boldsymbol{S}_{\boldsymbol{\mathcal{M}}}^{\textit{sn}}=\int m_s C_{\textit{sn}}\;\boldsymbol{v}  d\boldsymbol{v}$} are vectors representing momentum sources arising respectively from plasma-plasma and plasma-neutral interactions, $\boldsymbol{q}_s$ is the heat flux density while $Q_s$ and $S_p^{\textit{sn}}=\int m_s C_{\textit{sn}}  \boldsymbol{v}^2/2 d\boldsymbol{v}$ are heating terms arising respectively from plasma-plasma and plasma-neutral collisions. 

We consider turbulent phenomena on scales much larger than the plasma Debye length $\lambda_D$, and much slower than the plasma frequency, $w_p$ \citep{Brenno25}. In this regime, the quasi-neutrality condition

\begin{equation}
    \nabla \cdot \boldsymbol{J}=0
    \label{eqn:chargeconservation}
\end{equation}

\noindent
is assumed, where $\boldsymbol{J}=\sum_sq_sn_s\boldsymbol{V}_s$ is the current density of the plasma. The system of equations (\ref{eqn:continuityonespecies})--(\ref{eqn:chargeconservation}) is closed by the following Maxwell's equations 

\begin{equation}
    \nabla \cdot \boldsymbol{B} = 0,
    \label{eqn:Gauss}
\end{equation}

\begin{equation}
    \frac{\partial \boldsymbol{B}}{\partial t} = - \nabla \times \boldsymbol{E},
    \label{eqn:faraday}
\end{equation}

\begin{equation}
    \mu_0 \boldsymbol{J} = \nabla \times \boldsymbol{B}.
    \label{eqn:ampere}
\end{equation}

In addition, Braginskii's closure specifies the friction and heating terms present in equations (\ref{eqn:momentumonespecies}) and (\ref{eqn:energyonespecies}) \citep{Braginskii65}; that is

\begin{equation}
    R_{\parallel e} = - R_{\parallel i} = -0.71 n_e \boldsymbol{b}\cdot \nabla T_e + en_e\frac{J_\parallel}{\sigma_\parallel},
    \label{eqn:collisionmomentumsource}
\end{equation}

\begin{equation}
    Q_i = -3 \frac{m_e}{m_i}\frac{n_e}{\tau_e} (T_i-T_e)
    \label{eqn:ionclosure}
\end{equation}

\noindent
and

\begin{equation}
   Q_e = 3 \frac{m_e}{m_i} \frac{n_e}{\tau_e}(T_i-T_e) - \frac{0.71}{e}\boldsymbol{J}_\parallel \cdot \nabla T_e + \frac{J_\parallel^2}{\sigma_\parallel},
    \label{eqn:closure}
\end{equation}

\noindent
where $\tau_e$ is the electron collision time and $\sigma_\parallel$ is the parallel electrical conductivity. Additionally, the closure specifies the heat flux $\boldsymbol{q}_s$ and stress tensor $\boldsymbol{\pi}_s$. Their expressions, omitted here for shortness, can be found in \citep{Braginskii65}. It is important to note that they contain conductive and viscous effects and, depending on the parallel conductivity $\chi_{\parallel s} \propto T_s^{5/2}$ and viscosity $\eta_{\parallel s} \propto T_s^{5/2}$, they have a strong temperature dependence \citep{Braginskii65}. In particular, Braginskii's closure considers a highly collisional regime, adequate for the boundary region of the plasma. An important assumption of the closure is that Coulomb collisions are elastic and conserve mass, momentum and energy (while from the plasma perspective plasma-neutral collisions constitute sources or sinks). As a result, upon summation over the plasma species, the Coulomb collision terms vanish. In particular, the conservation law for the total energy yields:

\begin{gather}
\hspace{-1em}%
\begin{aligned}
\sum_s \left[ \frac{\partial \mathcal{H}_s}{\partial t}
+ \nabla \cdot  
    (\mathcal{H}_s \boldsymbol{V}_s
    + \boldsymbol{P}_s \cdot \boldsymbol{V}_s
    + \boldsymbol{q}_s)
  \right]
&= \sum_s \left(
  \frac{\mathcal{H}_s}{n_s} S_n
  + S_p^s
  + S_p^{sn}
\right) + \boldsymbol{J}\cdot \boldsymbol{E},
\end{aligned}
\label{eqn:energy}
\end{gather}

\noindent
where $\boldsymbol{J}\cdot \boldsymbol{E}$ represents the exchange of energy between the plasma and the electromagnetic field. Indeed, the same term can be found through Poynting's theorem in the energy of the electromagnetic field \citep{Brenno25}

\begin{equation}
     \frac{\partial \mathcal{H}_{EM}}{\partial t} +\nabla \cdot \boldsymbol{S} = -\boldsymbol{J}\cdot \boldsymbol{E} 
    \label{eqn:Poynting}
\end{equation}

\noindent
where $\boldsymbol{S}=\boldsymbol{E}\times \boldsymbol{B}/\mu_0$ is Poynting's vector and $\mathcal{H}_{EM}=(\varepsilon_0E^2 + B^2/\mu_0)/2$ is the electromagnetic energy density with $\varepsilon_0$ and $\mu_0$ the free-space permitivity and permeability. Nevertheless, for simplicity we do not develop further and preserve the $\boldsymbol{J}\cdot \boldsymbol{E}$ term in equation (\ref{eqn:energy}) as a source.

Similar conservation laws can be obtained for mass and momentum by summing equations (\ref{eqn:continuityonespecies}) and (\ref{eqn:momentumonespecies}) over the different species, while charge conservation is given directly by (\ref{eqn:chargeconservation}). Conservation laws like (\ref{eqn:energy}) are written in flux form as:

\begin{equation}
    \frac{\partial A}{\partial t} +\nabla \cdot\boldsymbol{\Gamma}_A = S_A
\end{equation}

\noindent
with $\boldsymbol{\Gamma}_A$ and $S_A$ the flux and the sources or sinks of the quantity $A$ (in our case $\mathcal{H}$). Solid surfaces act as a sink of particles and energy for the plasma. Hence, if boundary conditions at the vessel walls are not considered, the energy of the plasma is preserved except for the presence of external sources and interactions with neutral species.

\section{Drift-reduced Braginskii equations} \label{sec:ideal}

The drift approximation assumes that the turbulent perpendicular scale lengths are larger than the sound Larmor radii $\rho_s$ and the turbulent timescales of interest are larger than the ion cyclotron frequency $\Omega_{ci}$ \citep{Drake84,Zeiler97,Scott03,Simakov03,Brenno25}. This can be expressed in terms of the drift-expansion parameter $\epsilon = \partial_t/\Omega_{ci} \approx \rho_s^2/ L_\perp^2 \ll 1$ with $\Omega_{ci}=q_iB/m_i$, $\rho_s=c_s/\Omega_{ci}$, $c_s=\sqrt{T_e/m_i}$, while $L_\perp$ is the characteristic perpendicular length. In this limit, the motion of each species can be decomposed in a parallel and perpendicular contribution with respect to the magnetic field $\boldsymbol{B}$ as

\begin{equation}
    \boldsymbol{V}_s = \boldsymbol{V}_{\parallel s} + \boldsymbol{V}_{\perp s},
\end{equation}

\noindent
where the perpendicular component can be written in terms of different drift contributions obtained by crossing equation (\ref{eqn:momentumonespecies}) with the magnetic field $\boldsymbol{B}$ \citep{Drake84,Zeiler97,Simakov03}. To leading order in $\epsilon$, the drifts that are usually retained in numerical codes are the $\boldsymbol{E}\times \boldsymbol{B}$, diamagnetic, stress and polarization drifts defined as

\begin{equation}
    \boldsymbol{V}_E=\frac{\boldsymbol{b}\times \nabla \phi}{B}, 
    \label{eqn:ExB}
\end{equation}

\begin{equation}
    \boldsymbol{V}_{d s} =  \frac{\boldsymbol{b}\times \nabla p_s}{q_sn_sB},
    \label{eqn:diam}
\end{equation}

\begin{equation}
    \boldsymbol{V}_{\boldsymbol{\pi}s} = \frac{\boldsymbol{b}\times( \nabla \cdot \boldsymbol{\pi}_{s} )}{q_s n_s B} ,
    \label{eqn:stresstensordrit}
\end{equation}

\noindent
and

\begin{equation}
    \boldsymbol{V}_{\text{pol} \; s} = \frac{\boldsymbol{b}}{n_s\Omega_{cs}}\times \left[ \frac{\partial (n_s \boldsymbol{V}_s)}{\partial t} + \nabla \cdot (n_s \boldsymbol{V}_s \boldsymbol{V}_s)\right]
    \label{eqn:properpolarization}
\end{equation}

\noindent
where $\phi$ is the electric potential and $\boldsymbol{b}$ is the unit vector in the direction of the magnetic field \citep{Brenno25}. In the following, consistently with most fluid codes and following Zeiler \textit{et al.} (1997), we simplify $\boldsymbol{V}_{\text{pol} \;i}$ and include the stress tensor drift in its definition, that is

\begin{equation}
\begin{aligned}
    \boldsymbol{V}_{\text{pol} \; i} = &\frac{1}{n_i \Omega_{ci}} \left(
    \frac{\partial [n_i \boldsymbol{b} \times(\boldsymbol{V}_E +\boldsymbol{V}_{di})]}{\partial t} + (\boldsymbol{V}_E+\boldsymbol{V}_{di} + \boldsymbol{V}_{\parallel i}) \cdot \nabla[n_i \boldsymbol{b}\times (\boldsymbol{V}_E + \boldsymbol{V}_{di})] \right)\\
    &+ \boldsymbol{V}_{\boldsymbol{\pi}i}.
    \label{eqn:polarizationvel}
\end{aligned}
\end{equation}

\noindent
We note that, as pointed out in Halpern \textit{et al.} (2016), following a perturbative expansion to first order in the drift-expansion parameter $\epsilon$, the term representing the polarization advection $\boldsymbol{V}_{\text{pol}} \cdot \nabla [n_i \boldsymbol{b} \times(\boldsymbol{V}_E+\boldsymbol{V}_{di})]$, is neglected in equation \eqref{eqn:polarizationvel}. Additionally, by assuming the large aspect ratio limit ($R_0\gg a$ with $R_0$ and $a$ the major and minor radius of the tokamak), the term $\Big[n_i \boldsymbol{b}\times (\boldsymbol{V}_E+\boldsymbol{V}_{di} )\Big]\nabla  \cdot (\boldsymbol{V}_E+\boldsymbol{V}_{di} + \boldsymbol{V}_{\parallel i})$ is dropped. Finally, the term related to inhomogeneities in the direction of the magnetic field \mbox{$d \boldsymbol{b}/dt = \partial_t \boldsymbol{b} +(\boldsymbol{V}\cdot \nabla) \boldsymbol{b}$} is neglected \citep{Zeiler97, Halpern2015}.

As pointed out by several authors \citep{Zeiler97,Halpern2015,Brenno25}, the perturbative expansion performed to obtain the expression (\ref{eqn:polarizationvel}) for $\boldsymbol{V}_{\text{pol}}$ breaks momentum and energy conservation to order $\epsilon$. The $\mathcal{O} (\epsilon)$ spurious terms introduced by the expansion (as discussed in Section \ref{sec:reconstructing}), can be avoided by analytically inverting expression (\ref{eqn:properpolarization}) for the polarization velocity, as demonstrated in De Lucca \textit{et al.} (2026). However, this expression has not been implemented in numerical codes yet.

Finally, the perpendicular velocities for each species are given by:

\begin{align}
    &\textbf{V}_{\perp i} = \boldsymbol{V}_E + \boldsymbol{V}_{di} + \boldsymbol{V}_{\text{pol} \; i},\\
    &\textbf{V}_{\perp e} = \boldsymbol{V}_E + \boldsymbol{V}_{de},
\end{align}
\noindent
where $\boldsymbol{V}_{\text{pol} \; e}$ is neglected since it is proportional to the electron mass and $m_e/m_i \ll 1$. Introducing the drifts (\ref{eqn:ExB}), (\ref{eqn:diam}) and (\ref{eqn:polarizationvel}) in equations (\ref{eqn:continuityonespecies}) and (\ref{eqn:energyonespecies}) yields:

\begin{align}
     & \frac{\partial n_e}{\partial t} 
      + \nabla \cdot \left[ n_e (\boldsymbol{V}_E + \boldsymbol{V}_{de} + \boldsymbol{V}_{\parallel e}) \right]
      = S_n + S_n^{\textit{en}},
      \label{eqn:continuityelectron}
    \\[1em] 
     &\begin{aligned}
        \frac{\partial \mathcal{H}_e }{\partial t}
        + \nabla \cdot \left[ (\mathcal{H}_e\mathbb{I} 
        + \boldsymbol{P}_e )\cdot (\boldsymbol{V}_E + \boldsymbol{V}_{de} + \boldsymbol{V}_{\parallel e})
        + \boldsymbol{q}_e \right]
        = \\-e n_e (\boldsymbol{V}_{de} + \boldsymbol{V}_{\parallel e}) \cdot \boldsymbol{E}
        + Q_e + \boldsymbol{R}_{ e} \cdot \boldsymbol{V}_{ e} \\
        + \mathcal{H}_e S_n/n_e
        + S_{p}^e + S_p^{\textit{en}},
    \end{aligned}
    \label{eqn:energyelectron}
\end{align}

\begin{align}
     & \frac{\partial n_i}{\partial t} 
      + \nabla \cdot \left[ n_i (\boldsymbol{V}_E + \boldsymbol{V}_{di} + \boldsymbol{V}_{\parallel i} + \boldsymbol{V}_{\text{pol} \; i}) \right]
      = S_n + S_n^{\textit{in}},
      \label{eqn:continuityion}
      \\[1em]
     &\begin{aligned}
        \frac{\partial \mathcal{H}_i }{\partial t}
        + \nabla \cdot \left[ (\mathcal{H}_i\mathbb{I}+\boldsymbol{P}_i )\cdot (\boldsymbol{V}_E + \boldsymbol{V}_{di} + \boldsymbol{V}_{\parallel i} + \boldsymbol{V}_{\text{pol} \; i})
        + \boldsymbol{q}_i \right]
        = \\e n_i (\boldsymbol{V}_{di} + \boldsymbol{V}_{\parallel i} + \boldsymbol{V}_{\text{pol} \; i}) \cdot \boldsymbol{E}
        + Q_i + \boldsymbol{R}_{ i} \cdot \boldsymbol{V}_{i}\\
        + \mathcal{H}_i S_n/n_i
        + S_{p}^i + S_p^{\textit{in}},
    \end{aligned}
    \label{eqn:energyion}
\end{align}

\noindent
while the scalar equation for the parallel momentum is obtained by projecting (\ref{eqn:momentumonespecies}) along $\boldsymbol{b}$:

\begin{align}
     &\begin{aligned}
        \frac{\partial \mathcal{M}_{\parallel e}}{\partial t}
        + \nabla \cdot [(\boldsymbol{V}_E + \boldsymbol{V}_{de} + \boldsymbol{V}_{\parallel e}) \mathcal{M}_{\parallel e}] - \boldsymbol{\mathcal{M}}_e\cdot \frac{d \boldsymbol{b}}{dt}
       = \\- \boldsymbol{b} \cdot (\nabla \cdot \boldsymbol{P}_e)
        -e n_e E_\parallel
        + R_{\parallel e} + \mathcal{M}_{\parallel e} S_n/n_e
        + \boldsymbol{b}\cdot\boldsymbol{S}_{\boldsymbol{\mathcal{M}}}^{\textit{en}},
    \end{aligned}
    \label{eqn:momentumelectron}
    \\[1em]
    &\begin{aligned}
        \frac{\partial \mathcal{M}_{\parallel i}}{\partial t}
        + \nabla \cdot [(\boldsymbol{V}_E + \boldsymbol{V}_{di} + \boldsymbol{V}_{\parallel i}+ \boldsymbol{V}_{\text{pol} \; i}) \mathcal{M}_{\parallel i}] - \boldsymbol{\mathcal{M}}_i\cdot \frac{d \boldsymbol{b}}{dt}
       = \\- \boldsymbol{b}\cdot(\nabla \cdot \boldsymbol{P}_i)
        +e n_i E_\parallel
        + R_{\parallel i} + \mathcal{M}_{\parallel i} S_n/n_i
        + \boldsymbol{b}\cdot\boldsymbol{S}_{\boldsymbol{\mathcal{M}}}^{\textit{in}}.
    \end{aligned}
    \label{eqn:momentumion}
\end{align}

\noindent
with $\mathcal{M}_{\parallel s} = \boldsymbol{M}_{s} \cdot \boldsymbol{b}$ the parallel momentum of species $s$. Since the perpendicular component of the friction terms $\boldsymbol{R}_\perp$ is much smaller than the parallel one, it is assumed that $\boldsymbol{R}_s \cdot \boldsymbol{V}_s \approx R_{\parallel s}V_{\parallel s}$ in equations (\ref{eqn:energyelectron}) and (\ref{eqn:energyion}). While some drift-reduced Braginskii codes now include electromagnetic  fluctuations \citep{Giacomin22, Zholobenko24, Quadri24}, we focus here on the electrostatic, large-aspect ratio limit of the equations for the sake of simplicity, hence dropping $d\boldsymbol{b}/dt$ terms in the following. 

Equations (\ref{eqn:continuityelectron})--(\ref{eqn:momentumion}) are written in flux form and are usually cast into advective form by recursively substituting the equations of lower order moments into the higher order ones \citep{Zeiler97}, thus routinely solving for $V_{\parallel s}$ and $T_s$ instead of $\mathcal{M}_{\parallel s}$ and $\mathcal{H}_s$ in fluid codes \citep{Giacomin22,Zholobenko24,Ben18}. More precisely, by substituting the continuity equations (\ref{eqn:continuityelectron}) and (\ref{eqn:continuityion}) in the momentum equations (\ref{eqn:momentumelectron}) and (\ref{eqn:momentumion}) we obtain the parallel velocity equations:

\begin{equation}
\begin{aligned}
    &\frac{\partial V_{\parallel e}}{\partial t}
        + (\boldsymbol{V}_E + \boldsymbol{V}_{de} + \boldsymbol{V}_{\parallel e}) \cdot \nabla V_{\parallel e}+
        \frac{1}{m_e n_e} \boldsymbol{b}\cdot(\nabla \cdot \boldsymbol{P}_e) =\\
        &-\frac{e}{m_e} E_\parallel
        + \frac{1}{m_e n_e}R_{\parallel e}+ \frac{1}{m_e n_e}\boldsymbol{b}\cdot \boldsymbol{S}_{\boldsymbol{\mathcal{M}}}^{\textit{en}}-\frac{V_{\parallel e}}{n_e}S_n^{\textit{en}},
    \label{eqn:vparelectron}
\end{aligned}
\end{equation}

\begin{equation}
\begin{aligned}
        &\frac{\partial V_{\parallel i}}{\partial t}
        + (\boldsymbol{V}_E + \boldsymbol{V}_{di} + \boldsymbol{V}_{\parallel i}+ \boldsymbol{V}_{\text{pol} \; i})  \cdot \nabla V_{\parallel i}
       +\frac{1}{m_i n_i}\boldsymbol{b}\cdot(\nabla \cdot \boldsymbol{P}_i)
        = \\& \frac{e}{m_i} E_\parallel
        + \frac{1}{m_i n_i}R_{\parallel i} + \frac{1}{m_i n_i} \boldsymbol{b}\cdot\boldsymbol{S}_{\boldsymbol{\mathcal{M}}}^{\textit{in}} - \frac{V_{\parallel i}}{n_i}S_n^{\textit{in}}.
    \label{eqn:vparion}
\end{aligned}
\end{equation}

\noindent
Then, by using the relation (\ref{eqn:Brennosequation}) by De Lucca \textit{et al.} (2026)

\begin{equation}
    m_s n_s \left.\frac{d_s \boldsymbol{V}_s}{dt} \right|_\perp + \frac{S_n}{n_s}m_s n_s \boldsymbol{V}_{\perp s} - q_s n_s  \boldsymbol{B} \times (\boldsymbol{V}_{\perp s} - \boldsymbol{V}_{\text{pol} \;s}) = q_s n_s \boldsymbol{V}_s \times \boldsymbol{B},
    \label{eqn:Brennosequation}
\end{equation}

\noindent
along with the velocity equations (\ref{eqn:vparelectron}) and (\ref{eqn:vparion}), and the continuity equations (\ref{eqn:continuityelectron}) and (\ref{eqn:continuityion}), we transform the energy equations (\ref{eqn:energyelectron}) and (\ref{eqn:energyion}) into evolution equations for the temperature:

\begin{equation}
\begin{aligned}
     &\frac{\partial T_e }{\partial t}
        + (\boldsymbol{V}_E + \boldsymbol{V}_{de} + \boldsymbol{V}_{\parallel e}) \cdot\nabla  T_e
        +\frac{2}{3n_e}\boldsymbol{P}_e :\nabla (\boldsymbol{V}_E + \boldsymbol{V}_{de} + \boldsymbol{V}_{\parallel e})+ \frac{2}{3n_e}\nabla \cdot \boldsymbol{q}_e
        = \\
         &\frac{2}{3 n_e}Q_e + \frac{2}{3n_e}S_{p}^e + \frac{2}{3 n_e}(S_{ p }^{\textit{en}}- \boldsymbol{V}_{\parallel e} \cdot\boldsymbol{S}_{\boldsymbol{\mathcal{M}}}^{\textit{en}})-\left(\frac{T_e}{n_e} -\frac{m_e V_{\parallel e}^2}{3n_e} \right) S_n^{\textit{en}}\\&+ \frac{4 \mathcal{K}_\perp^e}{3n_e^2}(S_n+S_n^{\textit{en}})\hspace{26em}
    \label{eqn:temperatureelectron}
\end{aligned}
\end{equation}

\begin{equation}
\begin{aligned}
        &\frac{\partial T_i }{\partial t}
        + (\boldsymbol{V}_E + \boldsymbol{V}_{di} + \boldsymbol{V}_{\parallel i} + \boldsymbol{V}_{\text{pol} \; i}) \cdot \nabla T_i \\
        &+ \frac{2}{3 n_i} \boldsymbol{P}_i : \nabla (\boldsymbol{V}_E + \boldsymbol{V}_{di} + \boldsymbol{V}_{\parallel i} + \boldsymbol{V}_{\text{pol} \; i})
        + \frac{2}{3 n_i}\nabla \cdot \boldsymbol{q}_i
        = \\ &+\frac{2}{3 n_i} Q_i + \frac{2}{3 n_i}S_{p}^i + \frac{2}{3n}(S_{ p}^{\textit{in}} - \boldsymbol{V}_{\parallel i} \cdot \boldsymbol{S}_{\boldsymbol{\mathcal{M}}}^{\textit{in}}) - \left(\frac{T_i}{n_i} -\frac{m_i V_{\parallel i}^2}{3n_i} \right) S_n^{\textit{in}}\\
        &+ \frac{4 \mathcal{K}_\perp^i}{3n_i^2}(S_n+S_n^{\textit{in}}) .
    \label{eqn:temperatureion}
\end{aligned}
\end{equation}

\noindent
We note that the external density source $S_n$, present in the momentum and energy equations (\ref{eqn:energyelectron}), (\ref{eqn:energyion})--(\ref{eqn:momentumion}), does not appear in equations (\ref{eqn:vparelectron})--(\ref{eqn:temperatureion}), except for the usually neglected small terms proportional to the perpendicular kinetic energy \mbox{$\mathcal{K}_\perp^s=m_s n_s V_{\perp s}^2 /2$}. 

For numerical implementation, equations (\ref{eqn:continuityelectron}), (\ref{eqn:continuityion}), (\ref{eqn:vparelectron}), (\ref{eqn:vparion}), (\ref{eqn:temperatureelectron}) and (\ref{eqn:temperatureion}) are usually further simplified using a number of approximations that we list here.

\noindent
(i) The stress tensor terms, $\boldsymbol{\pi}_s : \nabla \boldsymbol{V}_{s}$, in the temperature equations (\ref{eqn:temperatureelectron}) and (\ref{eqn:temperatureion}) are assumed small and dropped, hence, the $\boldsymbol{P}_s:\nabla \boldsymbol{V}_{s}$ terms simplify to $p_s \nabla \cdot \boldsymbol{V}_s$ \citep{Zeiler97}.

\noindent
(ii) The polarization advection terms $\boldsymbol{V}_{\text{pol} \; i}\cdot \nabla V_{\parallel i}$ and $\boldsymbol{V}_{\text{pol} \; i}\cdot \nabla T_{i}$ are deemed small and dropped in (\ref{eqn:vparion}) and (\ref{eqn:temperatureion}), since the polarization drift is subleading in the $\epsilon$ ordering \citep{Zeiler97}. 

\noindent
(iii) Under the assumption of quasi-neutrality $n_i \approx n_e \approx n$, one solves only equation~(\ref{eqn:continuityelectron}) for the density and then the charge conservation or vorticity equation (\ref{eqn:chargeconservation}) follows from substracting  both continuity equations (\ref{eqn:continuityelectron}) and (\ref{eqn:continuityion}) giving

\begin{equation}
    \nabla \cdot (n \boldsymbol{V}_{\text{pol} \;i}) + \frac{1}{e} \nabla \cdot (\boldsymbol{J}_\parallel + \boldsymbol{J}_{d}) = 0
    \label{eqn:vorticity}
\end{equation}

\noindent
where the scalar vorticity $\Omega = \nabla \cdot \boldsymbol{\omega}$ and vector vorticity $\boldsymbol{\omega} = n\nabla_\perp \phi + \nabla_\perp p_i/e$ are often used to rewrite this equation and evolve $\Omega$ in time \citep{Halpern2015}. The time evolution of $\phi$ is then obtained by solving a Poisson equation

\begin{equation}
    \nabla \cdot(n\nabla_\perp \phi) = \Omega - \frac{\nabla_\perp p_i}{e}.
    \label{eqn:poisson}
\end{equation}

\noindent
(iv) Explicit calculation of the term $\frac{2}{3}T_i \nabla \cdot \boldsymbol{V}_{\text{pol} \;i}$ in equation (\ref{eqn:temperatureion}) is avoided using equation (\ref{eqn:vorticity}) to rewrite 

\begin{equation}
    \frac{2}{3}T_i \nabla \cdot \boldsymbol{V}_{\text{pol} \; i} = -\frac{2 T_i}{3n}\boldsymbol{V}_{\text{pol i}}\cdot  \nabla n - \frac{2T_i}{3en}\nabla \cdot(\boldsymbol{J}_d+\boldsymbol{J_\parallel}).
    \label{partialresult3}
\end{equation}

\noindent
where the polarization advection $-\frac{2 T_i}{3n}\boldsymbol{V}_{\text{pol i}}\cdot  \nabla n $ is neglected as in (ii).

\noindent
(v) The equation for $V_{\parallel i}$ (\ref{eqn:vparion}) is usually  reduced to a single-fluid equation by adding equations (\ref{eqn:vparelectron}) and (\ref{eqn:vparion}) and then neglecting electron inertia since $m_e/m_i \ll 1$, yielding

\begin{equation}
\begin{aligned}
     \frac{\partial V_{\parallel i}}{\partial t}
        + (\boldsymbol{V}_E + \boldsymbol{V}_{di} + \boldsymbol{V}_{\parallel i})  \cdot \nabla V_{\parallel i}
       = \\- \frac{1}{m_i n}\boldsymbol{b}\cdot[\nabla \cdot (\boldsymbol{P}_i+p_e\mathbb{I})]_
        + \frac{1}{m_i n_i} \boldsymbol{b}\cdot \boldsymbol{S}_{\boldsymbol{\mathcal{M}}}^{\textit{in}} - \frac{V_{\parallel i}}{n_i}S_n^{\textit{in}}
    \label{eqn:vparion2}
\end{aligned}
\end{equation}

\noindent
where the electron stress tensor $\boldsymbol{\pi}_e$ can also be dropped by neglecting electron inertia \citep{Zeiler97}. 

\noindent
(vi) The effect of Ohmic heating in the $Q_e$ term is neglected, dropping the last term $J_\parallel^2/\sigma_\parallel$ in equation (\ref{eqn:closure}) \citep{Zeiler97}.

\noindent
(vii) Since the typical numerical discretization strategy chosen to solve the system of equations is often based on finite differences methods, artificial diffusion or hyperdiffusion terms are added to all equations for numerical stability \citep{Giacomin22,Zholobenko24,Ben18,Dudson24}. Even codes using finite volumes sometimes make use of some form of diffusion \citep{Quadri24}. 

As a specific example, the drift-reduced Braginskii equations implemented in GBS \citep{Giacomin22} are

\begin{flalign}
      &\frac{\partial n}{\partial t} 
      + \nabla \cdot \left[ n (\boldsymbol{V}_E + \boldsymbol{V}_{de} + \boldsymbol{V}_{\parallel e}) \right]
      = S_{n} + S_n^{\textit{en}} +D_n \nabla_\perp^2n, &&
      \label{eqn:continuityelectronfinal}
\end{flalign}

\begin{flalign}
    &\nabla \cdot (n \boldsymbol{V}_{\text{pol} \;i}) + \frac{1}{e} \nabla \cdot (\boldsymbol{J}_\parallel + \boldsymbol{J}_{d}) +\frac{D_\Omega}{B^2} \nabla_\perp^2 \Omega= 0, &&
    \label{eqn:vorticityfinal}
\end{flalign}

\begin{flalign}
        &\frac{\partial V_{\parallel e}}{\partial t}
        + (\boldsymbol{V}_E + \boldsymbol{V}_{de} + \boldsymbol{V}_{\parallel e}) \cdot \nabla V_{\parallel e} + \frac{1}{m_e n}\boldsymbol{b}\cdot(\nabla \cdot \boldsymbol{P}_e)
       = && \nonumber \\
        &-\frac{e}{m_e} E_\parallel
        + \frac{1}{m_e n}R_{\parallel e}+ \frac{1}{m_e n}\boldsymbol{b}\cdot\boldsymbol{S}_{\boldsymbol{\mathcal{M}}}^{\textit{en}}-\frac{V_{\parallel e}}{n}S_n^{\textit{en}} + D_{V_{\parallel e}}\nabla_\perp ^2 V_{\parallel e} ,&&
    \label{eqn:vparelectronfinal}
\end{flalign}

\begin{flalign}
     &\frac{\partial V_{\parallel i}}{\partial t}
        + (\boldsymbol{V}_E + \boldsymbol{V}_{di} + \boldsymbol{V}_{\parallel i})  \cdot \nabla V_{\parallel i}+  \frac{1}{m_i n}\boldsymbol{b}\cdot[\nabla \cdot (\boldsymbol{P}_i+p_e\mathbb{I})]
       = && \nonumber \\
        & \frac{1}{m_i n} \boldsymbol{b}\cdot \boldsymbol{S}_{\boldsymbol{\mathcal{M}}}^{\textit{in}} - \frac{V_{\parallel i}}{n}S_n^{\textit{in}}+ D_{V_{\parallel i}}\nabla_\perp ^2 V_{\parallel i}, &&
    \label{eqn:vparionfinal}
\end{flalign}

\begin{flalign}
    &\frac{\partial T_e }{\partial t}
    + (\boldsymbol{V}_E + \boldsymbol{V}_{de} + \boldsymbol{V}_{\parallel e}) \cdot\nabla  T_e
    +\frac{2}{3n} p_e \nabla \cdot (\boldsymbol{V}_E + \boldsymbol{V}_{de} + \boldsymbol{V}_{\parallel e}) + \frac{2}{3n}\nabla \cdot \boldsymbol{q}_e
    = && \nonumber \\
    &+\frac{2}{3 n}\left( 3 \frac{m_e}{m_i} \frac{n}{\tau_e}(T_i-T_e) - \frac{0.71}{e}\boldsymbol{J}_\parallel \cdot \nabla T_e\right) + \frac{2}{3n}S_{p}^e+\frac{4 \mathcal{K}_\perp^e}{3n^2}(S_n+S_n^{\textit{en}}) && \nonumber \\
    &+ \frac{2}{3 n}(S_{ p }^{\textit{en}}- \boldsymbol{V}_{\parallel e} \cdot \boldsymbol{S}_{\boldsymbol{\mathcal{M}}}^{\textit{en}})-\left(\frac{T_e}{n} -\frac{m_e V_{\parallel e}^2}{3n} \right) S_n^{\textit{en}} + D_{T_e} \nabla_\perp^2T_e, &&
    \label{eqn:temperatureelectronfinal}
\end{flalign}

\begin{flalign}
    &\frac{\partial T_i }{\partial t}
    + (\boldsymbol{V}_E + \boldsymbol{V}_{di} + \boldsymbol{V}_{\parallel i}) \cdot \nabla T_i + \frac{2T_i}{3n} \nabla \cdot (n \boldsymbol{V}_{de}) && \nonumber\\&+ \frac{2}{3 n} p_i  \nabla \cdot (\boldsymbol{V}_E +  \boldsymbol{V}_{\parallel e}) -\frac{2T_i}{3en^2}\boldsymbol{J}_\parallel \cdot \nabla n+ \frac{2}{3 n}\nabla \cdot \boldsymbol{q}_i
    = && \nonumber \\ &-\frac{2}{3 n} \left[ 3 \frac{m_e}{m_i} \frac{n}{\tau_e}(T_i-T_e)\right] + \frac{2}{3 n}S_{p}^i+ \frac{4 \mathcal{K}_\perp^i}{3n^2}(S_n+S_n^{\textit{in}}) && \nonumber \\       &+ \frac{2}{3n}(S_{ p}^{\textit{in}} - \boldsymbol{V}_{\parallel i} \cdot \boldsymbol{S}_{\boldsymbol{\mathcal{M}}}^{\textit{in}}) - \left(\frac{T_i}{n} -\frac{m_i V_{\parallel i}^2}{3n} \right) S_n^{\textit{in}} + D_{T_i} \nabla_\perp^2 T_i. &&
    \label{eqn:temperatureionfinal}
\end{flalign}

\noindent
We note that these equations have only small differences with respect to most of the drift-reduced Braginskii codes \citep{Zholobenko24,Quadri24,Ben18,Dudson_2026}.

To leading order in $\epsilon$, equations (\ref{eqn:continuityelectronfinal}) and (\ref{eqn:vorticityfinal}) conserve mass and charge, but for the presence of artificial diffusion terms.
In the following, we analyze the effect of the perturbative expansion of $\boldsymbol{V}_{\text{pol i}}$, along with the rest of the approximations and artificial diffusion introduced in the set of equations solved by GBS, with specific focus on energy conservation.

\section{Reconstructing  and verifying energy conservation} \label{sec:reconstructing}

In the present section, we focus on the conservation of total energy \mbox{$\mathcal{H} = \sum_s (\mathcal{K}_s + \mathcal{U}_s)$}, that is the sum, over both species, of internal energy $\mathcal{U}_s=3/2 p_s$ and kinetic energy \mbox{$\mathcal{K}_s = \mathcal{K}_{\parallel s} + \mathcal{K}_{\perp s}$}, which in the drift-reduced limit is expressed as the sum of parallel kinetic energy, $\mathcal{K}_{\parallel s}=m_s n_s V_{\parallel s}^2/2$, and perpendicular kinetic energy, $\mathcal{K}_{\perp s}=m_s n_s V_{\perp s}^2/2$. We start from the advective-form equations (\ref{eqn:continuityelectronfinal})--(\ref{eqn:temperatureionfinal}) to rebuild the energy conservation law and compare it with equation (\ref{eqn:energy}). 

To derive a conservation equation for $\mathcal{H}$ from equations (\ref{eqn:continuityelectronfinal})--(\ref{eqn:temperatureionfinal}), we first add equations (\ref{eqn:continuityelectronfinal}) and (\ref{eqn:vorticityfinal}) to retrieve the ion continuity equation

\begin{equation}
    \frac{\partial n}{\partial t} 
      + \nabla \cdot \left[ n (\boldsymbol{V}_E + \boldsymbol{V}_{di} + \boldsymbol{V}_{\parallel i} + \boldsymbol{V}_{\text{pol} \; i}) \right]
      = S_{n} + S_n^{\textit{en}} +D_n \nabla_\perp^2n -\frac{ D_\Omega}{B^2} \nabla_\perp^2\Omega.
    \label{eqn:ioncontinuityfinal}
\end{equation}

\noindent
We note that, even if $n_e \approx n_i \approx n$ under the quasineutrality approximation, we use equation (\ref{eqn:ioncontinuityfinal}) instead of (\ref{eqn:continuityelectronfinal}) in the derivation of conservation laws in order to clearly differentiate the ion and electron transport terms. Comparing with equation (\ref{eqn:continuityion}), we see that equation (\ref{eqn:ioncontinuityfinal}) contains two non-conservative contributions originating from the numerical diffusion terms. Their combined effect is expected to be small relative to the other terms in equation \eqref{eqn:ioncontinuityfinal}.

We now follow previous work \citep{Zeiler97,Loizu2014}, to reconstruct the conservation laws for $\mathcal{U},\; \mathcal{K}_\parallel$ and $\mathcal{K}_\perp$. We start by considering the internal energy $\mathcal{U}=\mathcal{U}_e + \mathcal{U}_i$ and we expand

\begin{equation*}
    \frac{\partial}{\partial t} \left[ \frac{3}{2} (p_e + p_i) \right] = \frac{3}{2}T_e \frac{\partial n}{\partial t} + \frac{3}{2}n \frac{\partial T_e}{\partial t} +\frac{3}{2}T_i \frac{\partial n}{\partial t} + \frac{3}{2}n \frac{\partial T_i}{\partial t} 
\end{equation*}

\noindent
where the terms $\partial_t n$, $\partial_t T_e$, $\partial_t T_i$, can be evaluated by using the ion and electron continuity and temperature equations (\ref{eqn:continuityelectronfinal}), (\ref{eqn:temperatureelectronfinal}), (\ref{eqn:temperatureionfinal}) and (\ref{eqn:ioncontinuityfinal}), yielding:

\begin{equation}
\begin{aligned}
    \frac{\partial \mathcal{U}}{\partial t} = 
    -\nabla \cdot \left[ \frac{3}{2}p_i(\boldsymbol{V}_E + \boldsymbol{V}_{di}+\boldsymbol{V}_{\parallel i}) + \frac{3}{2}p_e(\boldsymbol{V}_{E}+\boldsymbol{V}_{de} + \boldsymbol{V}_{\parallel e}) + \boldsymbol{q}_i + \boldsymbol{q}_e 
    \right]
    \\ -p_i\nabla \cdot (\boldsymbol{V}_E + \boldsymbol{V}_{di}+\boldsymbol{V}_{\parallel i})- p_e \nabla \cdot (\boldsymbol{V}_E + \boldsymbol{V}_{de}+\boldsymbol{V}_{\parallel e})
    \\+ \frac{3}{2}  (T_e+T_i)S_n +2 \frac{\mathcal{K}_\perp^i}{n} (S_n+S_n^{\textit{in}})+ S_{p}^e +S_p^i -0.71 \frac{\boldsymbol{J}_\parallel}{e} \cdot \nabla T_e\\
    +(S_{p}^{\textit{en}}-\boldsymbol{V}_{\parallel e}\cdot \boldsymbol{S}_{\boldsymbol{\mathcal{M}}}^{\textit{en}}) +  \frac{m_e V_{\parallel e}^2}{2} S_n^{\textit{en}}\\
    +(S_{ p}^{\textit{in}}-\boldsymbol{V}_{\parallel i}\cdot \boldsymbol{S}_{\boldsymbol{\mathcal{M}}}^{\textit{in}}) +  \frac{m_i V_{\parallel i}^2}{2} S_n^{\textit{in}}\\
    + \frac{3}{2}(T_e+T_i)D_{n}\nabla_\perp^2n+\frac{3}{2}n(D_{T_e}\nabla_\perp^2T_e+D_{T_i} \nabla_\perp^2 T_i)\\
     +\frac{5T_i}{2}\left(-\nabla \cdot (n \boldsymbol{V}_{pol \; i}) - \frac{D_\Omega}{B^2} \nabla_\perp^2 \Omega \right).
    \label{eqn:internalenergy}
\end{aligned}
\end{equation}

\noindent
Similarly, for the parallel kinetic energy $\mathcal{K}_\parallel=\mathcal{K}_{\parallel e} +\mathcal{K}_{\parallel i}$, we expand

\begin{equation*}
    \frac{\partial (m_e n V_{\parallel e}^2/2 + m_i n V_{\parallel i}^2/2) }{\partial t} = m_e\frac{V_{\parallel e }^2}{2}\frac{\partial n}{\partial t} + m_e n V_{\parallel e}\frac{\partial V_{\parallel e}}{\partial t} + m_i \frac{V_{\parallel i }^2}{2}\frac{\partial n}{\partial t} + m_i n V_{\parallel i}\frac{\partial V_{\parallel i}}{\partial t}
\end{equation*}

\noindent
where we use the electron and ion continuity equations (\ref{eqn:continuityelectronfinal}) and (\ref{eqn:ioncontinuityfinal}) together with the parallel velocity equations (\ref{eqn:vparelectronfinal}) and (\ref{eqn:vparionfinal}) to obtain:

\begin{equation}
\begin{aligned}
    \frac{\partial \mathcal{K}_\parallel}{\partial t} = -\nabla \cdot \left[ \frac{m_enV_{\parallel e}^2}{2}(\boldsymbol{V}_E + \boldsymbol{V}_{de} + \boldsymbol{V}_{\parallel e}) + \frac{m_inV_{\parallel i}^2}{2}(\boldsymbol{V}_E +\boldsymbol{V}_{di} + \boldsymbol{V}_{\parallel i})   \right] \\
    -\boldsymbol{V}_{\parallel e}\cdot (\nabla \cdot \boldsymbol{P}_e)-\boldsymbol{V}_{\parallel i}\cdot (\nabla \cdot \boldsymbol{P}_i) \\
    +\boldsymbol{J}_\parallel \cdot \boldsymbol{E}-\frac{1}{en}J_\parallel R_{\parallel e}-\boldsymbol{V}_{\parallel i} \cdot (en \boldsymbol{E}_\parallel -R_{\parallel e}  +\nabla p_e) \\
    +\frac{m_e V_{\parallel e}^2 + m_i V_{\parallel i}^2}{2} S_n \\+ \boldsymbol{V}_{\parallel e} \cdot \boldsymbol{S}_{\mathcal{M}\parallel}^{\textit{en}} - \frac{m_e V_{\parallel e}^2}{2} S_n^{\textit{en}}+\boldsymbol{V}_{\parallel i} \cdot \boldsymbol{S}_{\mathcal{M}\parallel}^{\textit{in}} - \frac{m_i V_{\parallel i}^2}{2} S_n^{\textit{in}} \\
    +\frac{m_e V_{\parallel e}^2 + m_i V_{\parallel i}^2}{2}D_{n} \nabla_\perp^2n+ m_enV_{\parallel e} D_{V_{\parallel e}}\nabla_\perp^2 V_{\parallel e} + m_i nV_{\parallel i} D_{V_{\parallel i}}\nabla_\perp^2 V_{\parallel i} \\
   + \frac{m_iV_{\parallel i}^2}{2}\left(-\nabla \cdot (n \boldsymbol{V}_{pol} ) - \frac{D_\Omega}{B^2} \nabla_\perp^2 \Omega\right).
    \label{eqn:parallelkineticenergy}
\end{aligned}
\end{equation}

Regarding the perpendicular kinetic energy, the drift-reduced model does not have an explicit evolution of the electron energy since $\boldsymbol{V}_{\text{pol}\; e}$ is neglected. Thus, we only consider the ion perpendicular kinetic energy $\mathcal{K}_\perp=\mathcal{K}_{\perp}^i$. The conservation law is usually found by multiplying the vorticity equation with $\phi$ and rearranging as:

\begin{equation}
    \phi \left(\nabla \cdot \boldsymbol{J}  +e \frac{D_\Omega}{B^2} \nabla_\perp^2 \Omega \right)= \nabla\cdot(\phi \boldsymbol{J}) - (\boldsymbol{J}_\parallel + \boldsymbol{J}_d) \cdot \nabla \phi -\boldsymbol{J}_{\text{pol}}\cdot \nabla \phi  +e\phi \frac{D_\Omega}{B^2} \nabla_\perp^2 \Omega= 0,
\end{equation}

\noindent
where the polarization current only includes the ion contribution $\boldsymbol{J}_{\text{pol}}=e n \boldsymbol{V}_{\text{pol} \;i}$. After convenient rewriting and isolating the term involving the polarization current, we obtain

\begin{equation}
    \boldsymbol{J}_{\text{pol}}\cdot \frac{\boldsymbol{\omega}}{n} = \nabla\cdot(\phi \boldsymbol{J}) - (\boldsymbol{J}_\parallel + \boldsymbol{J}_d) \cdot \nabla \phi +\boldsymbol{J}_{\text{pol}} \cdot \frac{\nabla p_i}{en} +e\phi \frac{D_\Omega}{B^2} \nabla_\perp^2 \Omega.
    \label{eqn:middlestep}
\end{equation}

As shown in Appendix \ref{AppendixD}, the left-hand side of equation (\ref{eqn:middlestep}) can be expanded using the definition for $\boldsymbol{V}_{\text{pol} \;i}$ in equation (\ref{eqn:polarizationvel}) to yield:

\begin{equation}
\begin{aligned}
    \boldsymbol{J}_{\text{pol}}\cdot\frac{\boldsymbol{\omega}}{n} =-\frac{\partial \mathcal{K}_\perp}{\partial t} - \nabla \cdot [ (\boldsymbol{V}_E+\boldsymbol{V}_{di}+\boldsymbol{V}_{\parallel i})\mathcal{K}_\perp]\\
    -\frac{\mathcal{K}_\perp}{n}\left[-\nabla \cdot (n \boldsymbol{V}_{\text{pol}})-\frac{D_\Omega}{B^2} \nabla_\perp^2 \Omega +S_n + S_n^ {\textit{in}} +D_n \nabla_\perp^2 n)  \right]
    \\+ 2 \mathcal{K}_\perp \nabla \cdot (\boldsymbol{V}_E+\boldsymbol{V}_{di}+\boldsymbol{V}_{\parallel i})-(\boldsymbol{V}_E+\boldsymbol{V}_{di})\cdot (\nabla \cdot \boldsymbol{\pi}_i).
    \label{partialresult4}
\end{aligned}
\end{equation}

Substituting (\ref{partialresult4}) into \eqref{eqn:middlestep}, we find the conservation law for $\mathcal{K}_\perp$ to be 

\begin{equation}
\begin{aligned}
    \frac{\partial \mathcal{K}_\perp}{\partial t}  = - \nabla \cdot \left[\mathcal{K}_\perp(\boldsymbol{V}_E + \boldsymbol{V}_{di} +\boldsymbol{V}_{\parallel i})\right]\\
    -\frac{\mathcal{K}_\perp}{n}(S_n+S_n^{\textit{in}})\\
    -(\boldsymbol{V}_E + \boldsymbol{V}_{di}) \cdot (\nabla \cdot \boldsymbol{\pi}_i)
    -\nabla \cdot (\phi \boldsymbol{J}) -(\boldsymbol{J}_\parallel + \boldsymbol{J}_d )\cdot \boldsymbol{E} -\nabla p_i\cdot \boldsymbol{V}_{\text{pol}}\\
    -\frac{\mathcal{K}_\perp}{n}D_{n} \nabla_\perp^2n-\frac{e\phi }{B^2 }D_\Omega \nabla_\perp^2 \Omega\\
    +\frac{\mathcal{K}_\perp}{n}\left(\nabla \cdot (n \boldsymbol{V}_{\text{pol}} ) + \frac{D_\Omega}{B^2} \nabla_\perp^2 \Omega\right)\\
    +2\mathcal{K}_\perp \nabla \cdot (\boldsymbol{V}_E+\boldsymbol{V}_{di}+\boldsymbol{V}_{\parallel i}).
    \label{eqn:perpkineticenergy}
\end{aligned}
\end{equation}

Adding equations (\ref{eqn:internalenergy}), (\ref{eqn:parallelkineticenergy}) and (\ref{eqn:perpkineticenergy}) we find the evolution equation for the total energy $\mathcal{H}$:

\begin{equation}
\begin{aligned}
    \frac{\partial \mathcal{H}}{\partial t} \\ +\nabla \cdot \Big[ (\mathcal{H}_{\parallel e} + p_e)(\boldsymbol{V}_E + \boldsymbol{V}_{de} + \boldsymbol{V}_{\parallel e}) + (\mathcal{H}_{ i} + p_i)(\boldsymbol{V}_E +\boldsymbol{V}_{di} + \boldsymbol{V}_{\parallel i}) +p_i \boldsymbol{V}_{\text{pol i}}\Big] \\
    +\nabla \cdot \Big[\boldsymbol{q}_e + \boldsymbol{q}_i+
    \boldsymbol{\pi}_e\cdot\boldsymbol{V}_{\parallel e}+\boldsymbol{\pi}_i\cdot(\boldsymbol{V}_E +\boldsymbol{V}_{di} + \boldsymbol{V}_{\parallel i}) \Big]\\
    = \frac{\mathcal{H}}{n}S_n + S_{p}^e +S_p^i + S_{ p}^{\textit{en}} + S_{ p}^{\textit{in}} + \boldsymbol{J}\cdot \boldsymbol{E}\\
    +L_O + L_{\mathcal{M}} + L_\pi + L_{\text{pol}}+L_C + L_{D}
    \label{eqn:totalenergyreconstructed}
\end{aligned}
\end{equation}

\noindent
where we define $\mathcal{H}_{\parallel s}=\mathcal{U}_s+\mathcal{K}_{\parallel s}$ and the terms that violate energy conservation:

\begin{equation}
    L_O = -\frac{J_\parallel^2}{\sigma_\parallel},
    \label{eqn:ohmicsink}
\end{equation}

\begin{equation}
    L_{\mathcal{M}}=-\boldsymbol{V}_{\parallel i}\cdot (en \boldsymbol{E}_\parallel - \boldsymbol{R}_{\parallel e}+ \nabla p_e),
    \label{eqn:electroninertiasink}
\end{equation}

\begin{equation}
    L_\pi = \boldsymbol{\pi}_e : \nabla \boldsymbol{V}_{\parallel e}+\boldsymbol{\pi}_i : \nabla (\boldsymbol{V}_{E}+\boldsymbol{V}_{d i}+\boldsymbol{V}_{\parallel i}),
    \label{eqn:stresssink}
\end{equation}

\begin{equation}
    L_{\text{pol}} =- \frac{\mathcal{H}_{\parallel i}-\mathcal{K}_\perp}{n}\nabla\cdot (n \boldsymbol{V}_{\text{pol i}}) -T_i \boldsymbol{V}_{\text{pol i}}\cdot \nabla n,
    \label{eqn:polarizationsink}
\end{equation}

\begin{equation}
    L_C = 2\mathcal{K}_\perp \nabla \cdot (\boldsymbol{V}_E+\boldsymbol{V}_{di}+\boldsymbol{V}_{\parallel i})
    \label{eqn:curvaturesink}
\end{equation}

and

\begin{equation}
\begin{aligned}
    L_{D} = \frac{\mathcal{H}_{\parallel e} + \mathcal{H}_{\parallel i}-\mathcal{K}_\perp}{n}D_{n}\nabla_\perp^2n +\frac{3}{2}n(D_{T_e}\nabla_\perp^2T_e+D_{T_i} \nabla_\perp^2 T_i)\\+m_enV_{\parallel e} D_{V_{\parallel e}}\nabla_\perp^2 V_{\parallel e} + m_i nV_{\parallel i} D_{V_{\parallel i}}\nabla_\perp^2 V_{\parallel i}-\left(\frac{\mathcal{H}_{\parallel i}+p_i-\mathcal{K}_\perp}{n}\right) \frac{D_{\Omega}}{B^2}\nabla_\perp^2 \Omega.
    \label{eqn:diffusionsink}
\end{aligned}
\end{equation}

Comparing equation (\ref{eqn:totalenergyreconstructed}) with (\ref{eqn:energy}), we observe that the divergence term as well as the source terms in (\ref{eqn:totalenergyreconstructed}) coincide with the ones in (\ref{eqn:energy}) except for $p_i \boldsymbol{V}_{\text{pol i}}$ which should be $(\mathcal{H}_i +p_i) \boldsymbol{V}_{\text{pol i}}$. In contrast to (\ref{eqn:energy}), we report a total of six terms of different magnitudes which break energy conservation. First and possibly largest is $L_O$, due to the neglected Ohmic heating term and estimated to be of order $\mathcal{O}\Big([V_{\parallel e}/c_{s0}]^2\Big)\approx \mathcal{O}(1)$. Second, the finite electron inertia term $L_{\mathcal{M}}$ is smaller, of order $\mathcal{O}(m_e/m_i)$. Third, the term associated to stress tensor heating $L_\pi$ is of order $\mathcal{O}(c_{s0}\tau_{s}\sqrt{m_s/m_i}  /R_0)$. This term scales favourably with machine size and is small for the electrons but around $\mathcal{O}(1)$ for the ions, although neoclassical corrections of the viscosity would make this spurious term smaller also for the ions \citep{Rozhansky09}. Fourth, the energy sink due to the neglected polarization advection terms $L_{\text{pol}}$ is of order $\mathcal{O}(\epsilon)$. While the second term in $L_{\text{pol}}$, proportional to $\mathcal{K}_\perp$, is already identified by other authors \citep{Halpern23,Brenno25}, here we show that, by neglecting polarization advection everywhere else in our equations, $L_{\text{pol}}$ contains additional contributions. Fifth, after neglecting curvature terms in the definition (\ref{eqn:polarizationvel}) of $\boldsymbol{V}_{\text{pol i}}$, we obtain the spurious term $L_C$ of order $\mathcal{O}(\kappa\epsilon)$, with $\kappa$ the magnitude of the curvature vector $\boldsymbol{\kappa}= \boldsymbol{b}\cdot \nabla \boldsymbol{b}$. Finally, the artificial diffusion terms introduced for numerical stability add up to $L_D$. This is an energy sink, which can be $\mathcal{O}(1)$ as we show in Section \ref{sec:simulationresults}. We note that, while $L_O$ and $L_{\mathcal{M}}$ do not scale with the system size or the magnetic field strength, $L_\pi$, $L_{\text{pol}}$ and $L_{C}$ scale favourably with one or the other. In the case of $L_{\text{pol}}$ and $L_{C}$, the drift-expansion parameter decreases with $B$, as $\epsilon\propto \rho_s^2/L_\perp^2 \approx\rho_s^2 / L_p^2 \propto B_\varphi^{-0.59}$ where $B_\varphi$ is the toroidal magnetic field and we use Giacomin's scaling for the pressure gradient $L_p \propto B_\varphi^{-0.706}$ \citep{Giacomin_2021}. A similar result can be obtained if one approximates $L_\perp \approx \lambda_q \propto B_\varphi^{-0.78}$ as in the Eich scaling \citep{Eich11}, obtaining $\epsilon \propto B_\varphi^{-0.44}$.  A simple estimate for typical TCV parameters yields $m_e/m_i \approx 2.7\cdot 10^{-4}$, $\epsilon \approx 6.5 \cdot 10^{-2}$, $(V_{\parallel e}/c_{s0})^2 \approx1$ and finally $c_{s0}\tau_{i}/R_0 \approx 1$.

\section{Simulation results} \label{sec:simulationresults}

In this section, we present simulation results that allow us to analyse the conservation law formulated in Section \ref{sec:reconstructing}. Two simulations are carried out, the first is based on the standard GBS model and follows the energy conservation law (\ref{eqn:totalenergyreconstructed}) (standard model), while for the second one, $L_O$ and $L_\mathcal{M}$ are removed by dropping approximations (v) and (vi) in Section \ref{sec:ideal} (improved model). We focus on the $L_O$ and $L_\mathcal{M}$ terms since they do not scale with the magnetic field or the machine size, as explained in Section \ref{sec:reconstructing}. Therefore, the results of a reduced-size TCV simulation should remain relevant for simulations of other devices. 

The initial condition for both simulations is provided by a snapshot from a converged, reversed-field (ion diamagnetic drift pointed upwards from the core, away the X-point), half-sized TCV-X21 simulation. This L-mode TCV-X21 case is a lower single-null magnetic TCV configuration performed experimentally with an extensive experimental coverage for code validation \citep{Oliveira22}. To decrease the computational cost of the two simulations, we use a reduced mass ratio, $m_i/m_e=450$, and keep the viscosity and thermal conductivity coefficients constant and set them to $\eta_{\parallel e}=\eta_{\parallel i}=1$ and $\chi_{\parallel e}=\chi_{\parallel i}=2.5$ in normalized GBS units (see apendix \ref{AppendixA}) \citep{Giacomin22}. The normalized electrical conductivity is set to $\sigma_0=10$ and both simulations have the same density and temperature sources. Even if the neutral dynamics are included in GBS \citep{Giacomin22}, we neglect their presence in the present simulations. While diffusion coefficients are set to $D=16$ in the initial simulation used as a starting point, we reduce them to $D_n=D_{T_e}=D_{T_i}=D_{V_{\parallel i}}=7$ and $D_{V_{\parallel e}}=D_\Omega=12$ in both the standard and improved simulations, in order to reduce the energy sink $L_D$ as much as possible. Following a transient, both simulations stabilize close to a quasi-steady state. We focus our analysis on the final 10 $t_0$ of both simulations, with $t_0= R_0/c_{s0}$ the GBS normalized time unit and $R_0$ the major radius. 

\begin{figure} 
    \centering
    \includegraphics[width=1.0\textwidth]{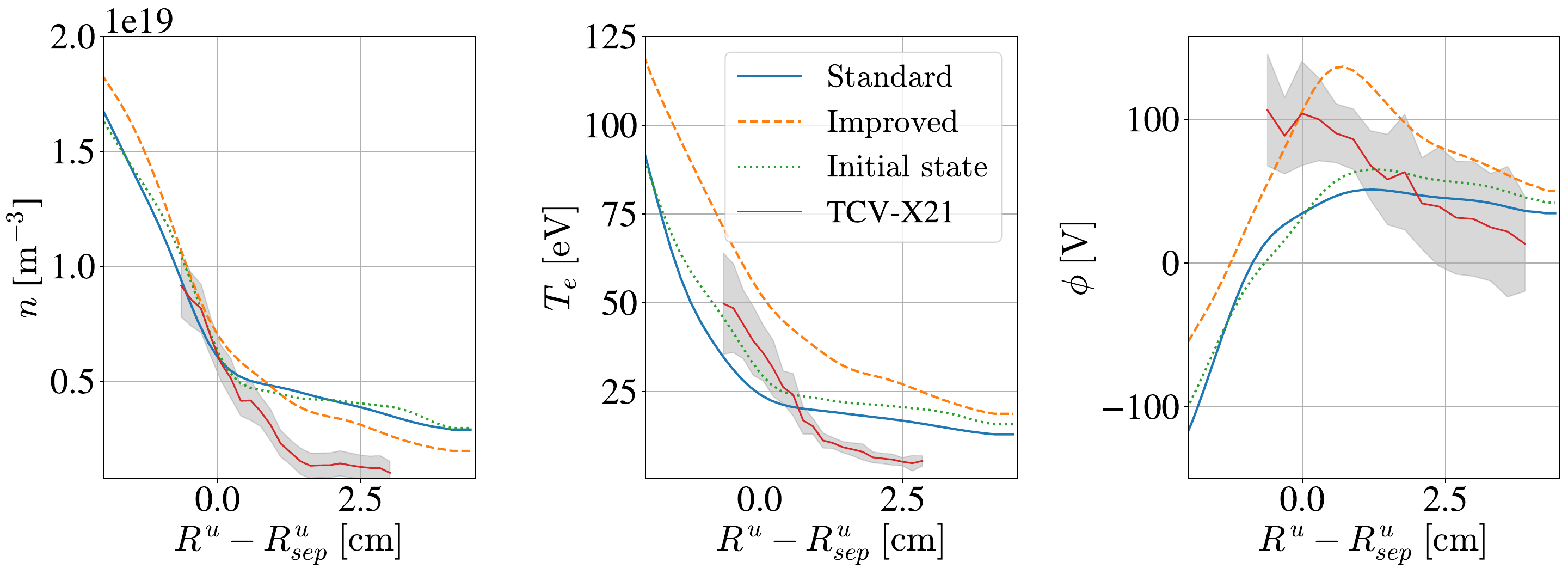}
    \caption{Density, temperature and electrostatic potential profiles at the upstream location compared with experimental measurements. The initial state is a converged simulation with $D=16$, standard and improved simulations are carried out with $D_n=D_{T_e}=D_{T_i}=D_{V_{\parallel i}}=7$ and $D_{V_{\parallel e}}=D_\Omega=12$.}
    \label{fig:upstream}
\end{figure}

The upstream profiles from the simulations are shown in figure \ref{fig:upstream} and compared to the TCV-X21 experimental data. Note that, here, the sources are not adjusted in order to match the upstream separatrix value of the density and temperature profiles with the experimental benchmark. Instead, we let the profiles evolve with the same sources of the original simulation and focus on the effect of removing $L_O$ and $L_\mathcal{M}$.
The density profile remains close to the initial state in both cases. The temperature and electrostatic potential decrease slightly for the standard case due to the reduction of artificial diffusion. On the other hand, the improved simulation shows an increase in both $T_e$ and $\phi$ with a steeper potential profile. This is expected since $L_O$ acts as an energy sink and its removal allows for higher temperatures. Indeed, the pressure profile steepens, with the pressure decay length decreasing from $L_p=11.89\,$mm in the standard case to $9.37\,$mm in the improved one. 

\begin{figure} 
    \centering
    \includegraphics[width=1.0\textwidth]{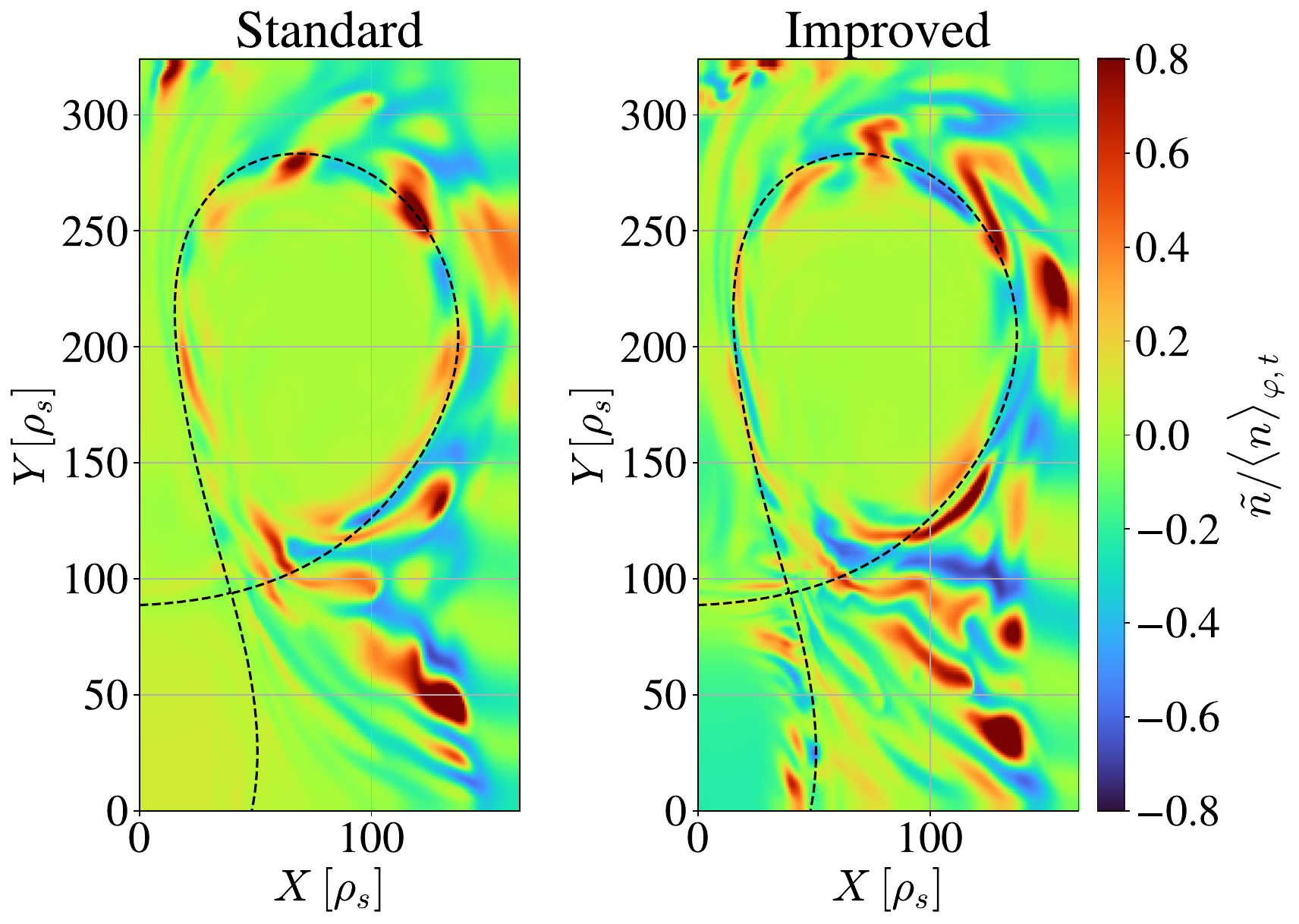}
    \caption{Comparison of normalized density fluctuations between the standard and improved simulations. A poloidal snapshot is considered.}
    \label{fig:Fluctuations}
\end{figure}

The density fluctuations $\tilde{n}=n-\langle n \rangle_{\varphi,t}$, normalized to the toroidal and time averaged density $\langle n \rangle_{\varphi,t}$, is presented for the two simulations in figure \ref{fig:Fluctuations}. Both poloidal profiles are qualitatively similar, with fluctuations concentrated on the low-field side of the domain, consistently with the ballooning nature of turbulence. Nevertheless, the improved simulation shows larger fluctuation levels and larger amplitude structures, particularly in the leg region close to the bottom wall.

\begin{figure} 
    \centering
    \includegraphics[width=1.0\textwidth]{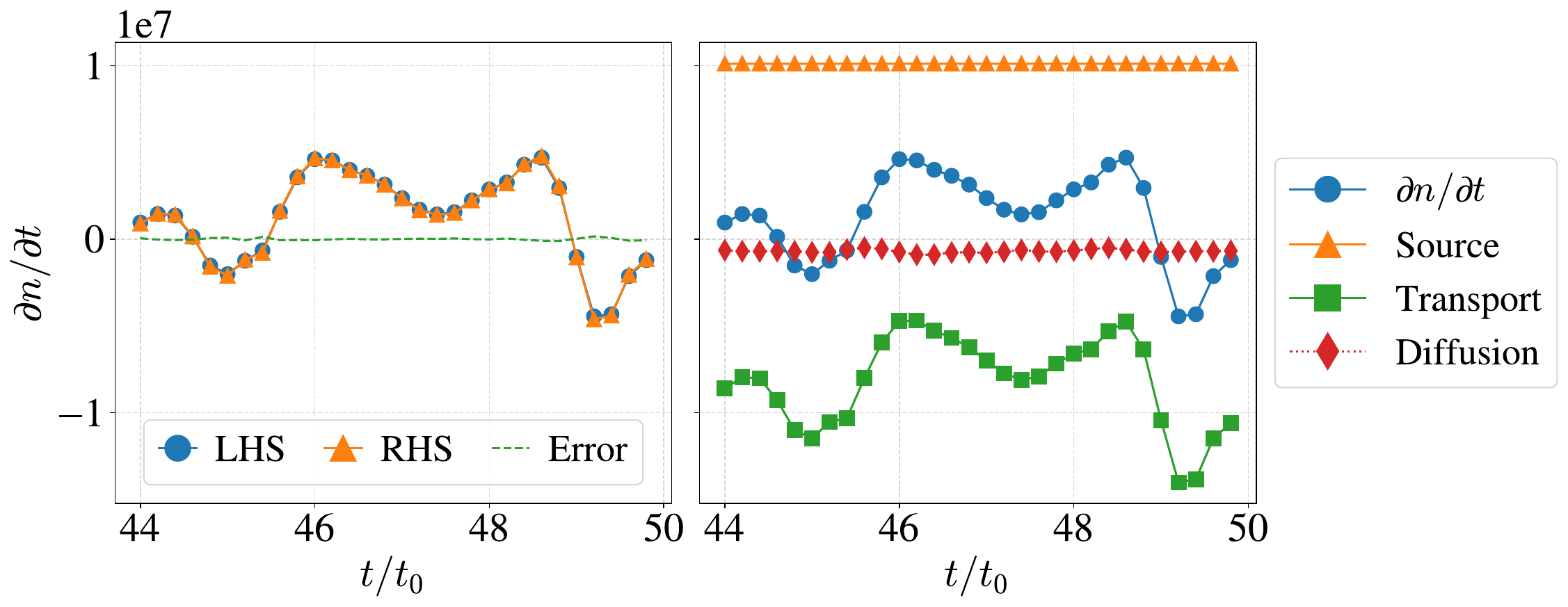}
    \caption{Verification of the continuity equation implemented in GBS (standard simulation).}
    \label{fig:continuity}
\end{figure}

Turning to the conservation laws, we exemplify and focus first on verifying the continuity equation (\ref{eqn:continuityelectronfinal}) implemented in GBS. The temporal derivative in the left-hand side is computed using centered differences and $\Delta t = 0.1 t_0$ spaced snapshots while the right-hand side is evaluated using the same numerical algorithms as the ones implemented in GBS \citep{Giacomin22}. Every term of the equation is then integrated over the core volume, where the sources are present. The left panel of figure \ref{fig:continuity} shows a very good agreement between both sides of the equation with a small residual. The right panel shows the different components of the right-hand side and their relative importance. We note that with $D_n=7$, the effect of artificial diffusion is small compared to the transport terms. 

\begin{figure} 
    \centering
    \includegraphics[width=1.0\textwidth]{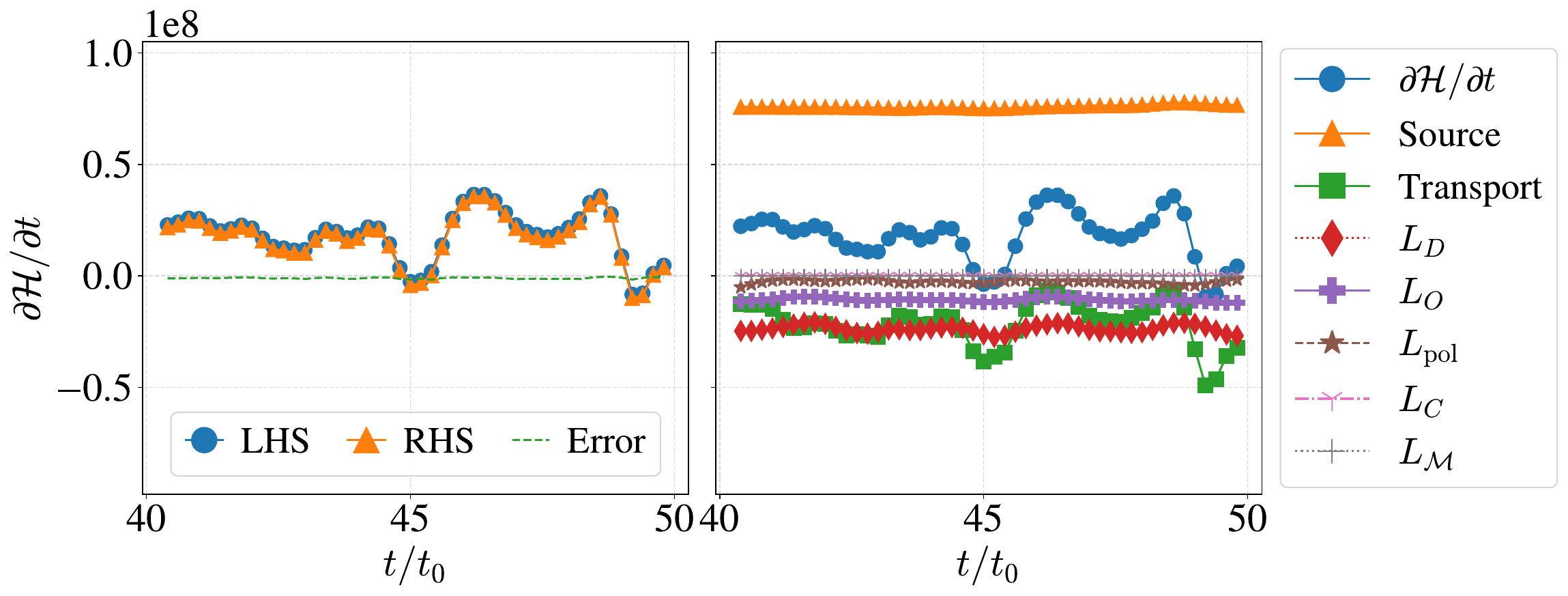}
    \caption{Verification of the energy equation previously implemented in GBS (standard simulation). The last two terms $L_C$ and $L_{\mathcal{M}}$ in the legend are very small and overlap on the plot.}
    \label{fig:energystandard}
\end{figure}

\begin{figure} 
    \centering
    \includegraphics[width=1.0\textwidth]{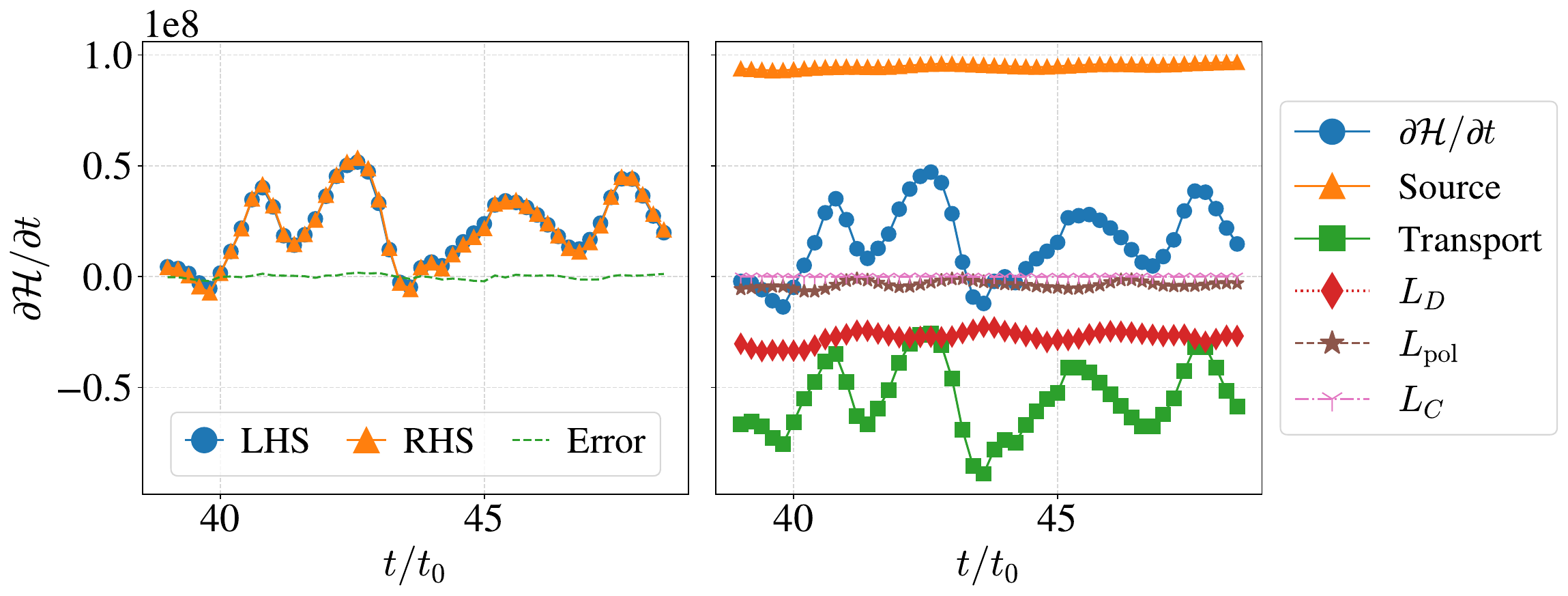}
    \caption{Verification of the energy equation currently implemented in GBS (improved simulation).}
    \label{fig:energyimproved}
\end{figure}

Figure \ref{fig:energystandard} shows the verification of the energy conservation law (\ref{eqn:totalenergyreconstructed}) for the standard simulation. On the left panel, the left- and right-hand sides of equation (\ref{eqn:totalenergyreconstructed}) are shown, also displaying a small residual. On the right panel of figure \ref{fig:energystandard}, we study the relative importance of the terms that compose the right-hand side of equation (\ref{eqn:totalenergyreconstructed}). The $L_\pi$ term is very small due to using a constant viscosity and its effect is added to the transport terms to simplify the analysis. The other energy conservation breaking terms have the expected relative magnitudes, notably with $L_D$ and $L_O$ representing the largest spurious contributions. We note that $L_D$ is most often comparable or larger than the transport terms, resulting in a considerable energy sink. In contrast, the results for the improved simulation in figure \ref{fig:energyimproved} show a similar magnitude of the left-hand side of the energy conservation law and of $L_D$ but with increased transport levels. As expected, the terms $L_{\text{pol}}$ and $L_C$, remain small and of similar magnitude for both simulations. We also remark that the energy source is larger in the improved simulation, because of the higher core temperatures. In fact, the density source acts as a power source in equation (\ref{eqn:totalenergyreconstructed}), proportional to the plasma temperature, $3/2 (T_i +T_e) S_n$.

In order to look quantitatively at the energy balance in the core region displayed in figures \ref{fig:energystandard} and \ref{fig:energyimproved}, we rewrite equation (\ref{eqn:totalenergyreconstructed}) as follows

\begin{equation}
    \frac{\Big( \frac{\partial \mathcal{H}}{\partial t} + \nabla\cdot \Gamma_{\mathcal{H}} - L_D - L_O -L_{\mathcal{M}} - L_{\text{pol}} - L_C \Big) }{\sum S} = 1,
    \label{eqn:percentages}
\end{equation}

\noindent
where we normalize the left-hand side $\partial \mathcal{H}/\partial t$, the transport of energy $\nabla \cdot \Gamma_{\mathcal{H}}$ and the spurious terms $L_D,\; L_O, \; L_{\mathcal{M}}, \; L_{\text{pol}}$ and $L_C$ by the sum of all sources in equation (\ref{eqn:totalenergyreconstructed}), $\sum S = \mathcal{H}S_n/ n + S_p^e + S_p^i + \boldsymbol{J}\cdot \boldsymbol{E}$. In table \ref{table:energybalancecore}, we list the relative contribution of each term in equation (\ref{eqn:percentages}) in the core region of both simulations. We see that both the standard and improved simulation are close to steady-state with the left-hand side $\partial \mathcal{H}/\partial t$ accounting for, approximately $20-23$\% of the power balance. Focusing on the other terms, while in a perfect steady-state and energy conserving simulation the transport term $\nabla \cdot \Gamma_{\mathcal{H}}$, should perfectly balance the sources, accounting for $100$\%, for the standard simulation this represents a contribution of only $28.14$\%. In addition to the finite value of $\partial \mathcal{H}/\partial t$ (the simulation is not in perfect steady state) the presence of spurious sinks is contributing to balance the sources and add up to a total $48.36$\%, with the major contributions being the one of artificial diffusion, $L_D$, and Ohmic heating, $L_O$. In practice, these sinks reduce the power source in the core, leading to lower temperature profiles for the standard simulation as displayed on figure \ref{fig:upstream}. In contrast, the transport term makes up $50.95$\% of the energy conservation contributions in the improved simulation. This is mainly the consequence of removing $L_O$ ($\partial \mathcal{H}/\partial t$ and $L_D$ are also slightly smaller than in the standard simulation). The terms $L_{\text{pol}}$ and $L_C$ are added together given their small contributions and show a very similar contribution in both simulations. The $L_{\mathcal{M}}$ spurious term represents a small correction and only slightly affects the simulation results as we tested with a separate simulation where only $L_O$ is corrected. We conclude that, the system reaches a different steady-state where transport is underestimated in the standard simulation. This is consistent with the larger fluctuation levels shown in figure \ref{fig:Fluctuations} once $L_O$ and $L_{\mathcal{M}}$ are corrected in the improved simulation.

\begin{table}
\centering
\begin{tabular}{c|c|c|c|c|c|c}
Model & $\partial \mathcal{H}/\partial t$ & $\nabla \cdot \Gamma_\mathcal{H}$ & \textbf{$-L_D$} & \textbf{$-L_O$} & $-L_\mathcal{M}$ & $-(L_{\text{pol}}+L_C)$ \\
\hline
\textbf{Standard} & 23.5 \% & 28.14 \% & 30.91 \% & 13.94 \% & 0.02 \% & 3.49 \% \\
\textbf{Improved} & 20.3 \% & 50.95 \% & 25.45 \% & 0.0 \% & 0.0 \% & 3.3 \%\\
\hline
\end{tabular}
\caption{Energy balance in the core region for the last $10 \;t_0$ of both simulations. All quantities are normalized to the total sum of sources $\sum S$. The improved simulation shows larger transport after removing the spurious terms $L_O$ and $L_{\mathcal{M}}$.}
\label{table:energybalancecore}
\end{table}

\begin{figure} 
    \centering
    \includegraphics[width=0.92\textwidth]{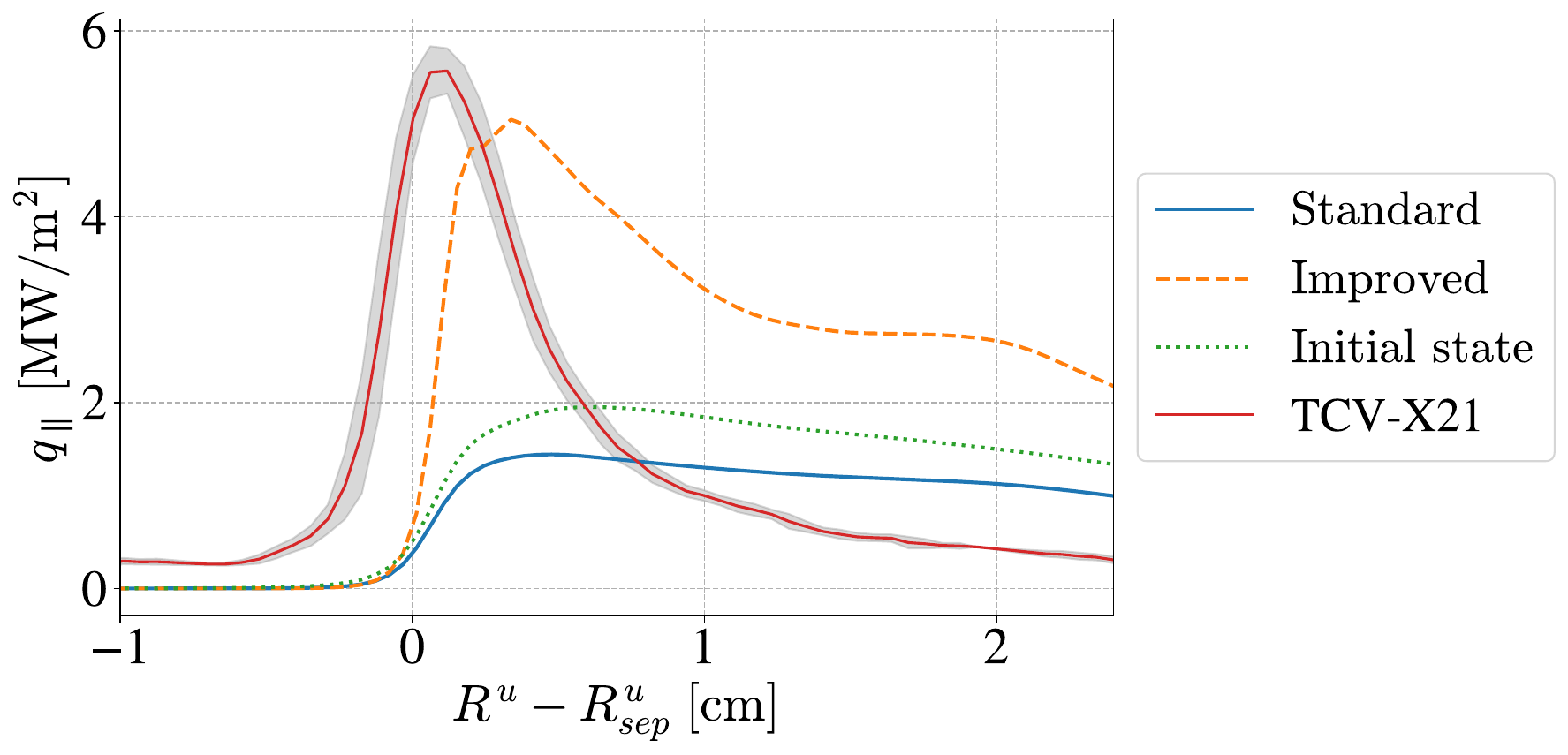}
    \caption{Target heat flux profile comparison between the simulations (half-size TCV) and the TCV-X21 experiments.}
    \label{fig:heatflux}
\end{figure}

With the caveats of a validation exercise carried out with reduced size simulations and absence of adaptive sources, we compare our estimate of the target heat flux to the experimental TCV-X21 database \citep{Oliveira22}. The heat flux is defined consistently with Oliveira \textit{et al.} (2022) as $q_{\parallel} = (m_e n V_{\parallel e}^2/2+5 p_e/2)V_{\parallel e} +(m_i n V_{\parallel i}^2/2+5 p_i/2)V_{\parallel i}$ (given the small chosen values of $\chi_{\parallel i}$ and $\chi_{\parallel e}$, we neglect conduction). In figure \ref{fig:heatflux}, we show the initial estimate of $q_\parallel$ obtained from the initial condition of the standard and improved simulations. Having reduced artificial diffusion, the heat flux from the standard simulation decreases, underestimating even more the experimental values, as noticed also in the full-size GBS simulation in Oliveira \textit{et al.} (2022). On the other hand, the improved simulation shows a significant increase in the estimated heat flux, which is now comparable to the experimental result and with similar peak magnitudes. A detailed analysis shows that this improvement is mostly associated with the removal of the Ohmic heating sink $L_O$, which has a large impact on the dynamics of the SOL region, as seen also in figure \ref{fig:Fluctuations}. Nevertheless, we note that the estimate of the heat flux decay length at the target $\lambda_q$, remains very similar to GBS values in Oliveira \textit{et al.} (2022) for the two simulations, with $16 \; \text{mm}$ for the standard simulation and $15.58 \; \text{mm}$ for the improved one, compared to the experimental value $4$ mm. This might be due to the the small and constant choice of electron heat conductivity $\chi_{\parallel e}$.

\begin{figure} 
    \centering
    \includegraphics[width=0.8\textwidth]{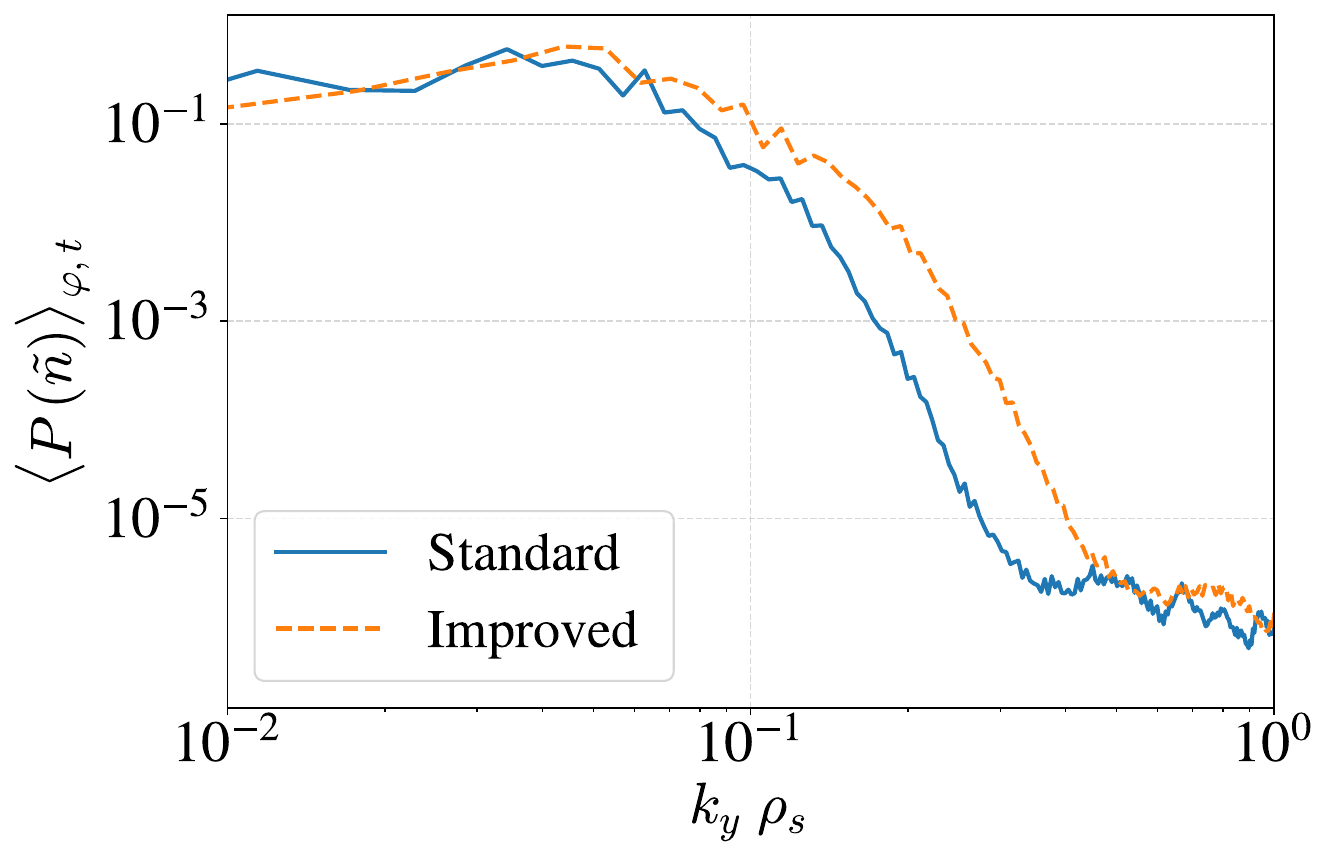}
    \caption{Power spectra of the density fluctuations of both simulations at the $\rho_\psi=0.98$ flux surface.}
    \label{fig:powerspectra}
\end{figure}

Finally, to further support our hypothesis that spurious terms reduce the strength of transport, we calculate the power spectrum of density fluctuations for both simulations. The results can be seen in figure \ref{fig:powerspectra}. As expected, due to the larger pressure gradient upstream, the power spectrum of the improved simulation shows that the fluctuations have larger amplitude for the same wavenumber, thus contributing to a stronger $E \times B$ transport.

\section{Conclusion}\label{sec:Conclusion}

The present paper reports on an analysis of the different approximations usually adopted for the implementation of the drift-reduced Braginskii model and their impact on energy conservation. As a practical example, we consider the GBS simulation code, allowing us to reconstruct the energy conservation law solved by the code and identify the effect of the different approximations. When the Ohmic heating and finite electron mass approximations, $L_O$ and $L_{\mathcal{M}}$, are avoided, the simulation shows an increased transport level and an improved estimate of the target heat flux with respect to the experimental results.

The plasma dynamics in the boundary results from the interplay between sources, sinks and transport terms. We show that, in the presence of non-physical energy sinks, energy transport is underestimated. For this reason, we expect convective transport to be overestimated in our simulation as a consequence of using small and constant conductivity coefficients. We believe this is the main reason behind the mismatch between the simulation and experimental profile shapes in figure \ref{fig:heatflux}.

An important energy sink results from artificial diffusion required by the numerical finite difference discretization scheme. This represents an effective sink of energy (denoted as $L_D$ here), and has a leading order effect on the solution. Since the equations are implemented in advective form, this effect results from the accumulation of several artificial diffusion terms. In fact, writing the system in flux form could help reduce the impact of artificial diffusion on the simulation results. Alternatively, a different discretization scheme, such as finite elements or finite volumes, could substantially reduce its impact. The artificial diffusion of the present scheme is added uniformly and in all directions, independently of the solution, so it acts as a persistent energy sink that biases the result. In contrast, the stabilization used by other methods, for example SUPG (Streamline-Upwind/Petrov-Galerkin) in finite elements \citep{Brooks} or flux-limited schemes in finite volumes \citep{LeVeque}, is proportional to the residual of the discretized equation, thus adding dissipation only where it is needed. In the SUPG case this dissipation is moreover confined to the streamwise direction, so it does not smear the solution across the flow.

We also quantify the well-known subleading terms resulting from neglecting polarization advection terms in the model equations (i.e., $L_{\text{pol}}$ and $L_C$), and show that they are small, as expected. However, we believe it would be of great interest to implement the expression of the vorticity equation provided by De Lucca \textit{et al.} (2026) and verify that the impact on the turbulent dynamics and fluctuating properties of the system are effectively small.

A natural extension of the present work would be to consider momentum conservation as well. From the present analysis, we expect the momentum conservation law to have an equivalent expression for the terms $L_\mathcal{M}$, $L_{\text{pol}}$, $L_C$ and $L_D$ with similar ordering. Furthermore, electromagnetic fluctuations and their approximations should be analysed. Similarly, the stress tensor terms $L_\pi$, estimated to be leading order, call for their accurate modeling. We expect that properly treating these terms may help current fluid codes to improve their agreement with experimental results and thus their predictive capabilities.

\section*{Acknowledgements}
The authors thank J. Mencke for insightful discussions about the different terms in the Boltzmann equation. This work has been carried out within the framework of the EUROfusion Consortium, partially funded by the European Union via the Euratom Research and Training Programme (Grant Agreement No 101052200 — EUROfusion). The Swiss contribution to this work has been funded in part by the Swiss State Secretariat for Education, Research and Innovation (SERI).
Views and opinions expressed are however those of the author(s) only and do not necessarily reflect those of the European Union, the European Commission or SERI. Neither the European Union nor the European Commission nor SERI can be held responsible for them. This work was supported in part by the Swiss National Science Foundation. The simulations were performed using computational resources provided by the LUMI supercomputer under the project ID EHPC-EXT-2024E02-100, the Pitagora supercomputer at CINECA under the project \textit{FUPA1LEGscan}, and by the ALPS-DAINT cluster at CSCS under Project ID lp71 \textit{Fluctuation and transport predictions in detached long-legged tokamak geometries}.

\section*{Declaration of interest}
The authors report no conflict of interest.
\section*{Data Availability Statement}
The data that supports the findings of this study is available from the corresponding author upon reasonable request.

\appendix

\section{Derivation of the $\mathcal{K}_\perp$ conservation law} \label{AppendixD}

We expand the term $\boldsymbol{J}_{pol}\cdot\boldsymbol{\omega}/n$ by introducing the definition of the polarization velocity given by (\ref{eqn:polarizationvel}):

\begin{equation}
\begin{aligned}
    &\boldsymbol{J}_{pol}\cdot\frac{\boldsymbol{\omega}}{n} =\left( \frac{m_i}{B} \frac{d [n\boldsymbol{b}\times(\boldsymbol{V}_E+\boldsymbol{V}_{di})]}{dt} + en\boldsymbol{V}_{\boldsymbol{\pi}_i}\right)\cdot \frac{\boldsymbol{\omega}}{n}.
    \label{eqn:appendparresult1}
\end{aligned}
\end{equation}

\noindent
We then integrate the first term in equation \eqref{eqn:appendparresult1} by parts. This yields

\begin{equation}
\begin{aligned}
    \frac{m_i }{B}   \frac{d (n \boldsymbol{b}\times[\boldsymbol{V}_E+\boldsymbol{V}_{di}])}{dt} =m_i \frac{d}{dt} \left( \frac{\boldsymbol{b}}{B} \times n[\boldsymbol{V}_E+\boldsymbol{V}_{di}]\right) - m_i \frac{d}{dt} \left( \frac{1}{B}\right) \boldsymbol{b}\times n (\boldsymbol{V}_E+\boldsymbol{V}_{di})\\
    = -m_i \frac{d}{dt} \left[ \frac{1}{B^2} \left(n \nabla_\perp \phi + \frac{\nabla_\perp p_i}{e}\right)\right] - m_i \frac{d}{dt} \left( \frac{1}{B}\right) \boldsymbol{b}\times n (\boldsymbol{V}_E+\boldsymbol{V}_{di}),
    \label{eqn:appendparresult2}
\end{aligned}
\end{equation}

\noindent
where we identify $\boldsymbol{\omega}/B^2$ inside the time derivative of the first term. The second term in (\ref{eqn:appendparresult2}) contains $d_t(1/B)$ and is present because in the definition (\ref{eqn:polarizationvel}) of $\boldsymbol{V}_{pol \; i}$, it is assumed $d_t \boldsymbol{b}\approx 0$.
For simplicity we drop it here to focus on the electrostatic and large aspect ratio limits (this term could be important in electromagnetic models). Re-injecting this result in (\ref{eqn:appendparresult1}) gives

\begin{equation}
\begin{aligned}
    \boldsymbol{J}_{pol}\cdot\frac{\boldsymbol{\omega}}{n} = -m_i  \frac{d }{dt} \left(\frac{\boldsymbol{\omega}}{B^2} \right)  \cdot \frac{\boldsymbol{\omega}}{n} +e \boldsymbol{V}_{\boldsymbol{\pi}_i}\cdot \boldsymbol{\omega}.
    \label{eqn:appendparresult3}
\end{aligned}
\end{equation}

Here, we recognize that $\mathcal{K}_\perp$ can be written in terms of the vector vorticity $\boldsymbol{\omega}$ as 

\begin{equation}
    \mathcal{K}_\perp=\frac{1}{2}\frac{m_i}{B^2}\frac{\boldsymbol{\omega}}{n}\cdot\boldsymbol{\omega},
    \label{eqn:perpkinenergy}
\end{equation}

\noindent
which allows us to rewrite the first term in (\ref{eqn:appendparresult3}) as 

\begin{equation}
\begin{aligned}
    m_i  \frac{d }{dt} \left(\frac{\boldsymbol{\omega}}{B^2} \right)  \cdot \frac{\boldsymbol{\omega}}{n} = \frac{1}{n} \frac{d}{dt}\left( \frac{\boldsymbol{\omega}\cdot \boldsymbol{\omega}}{2 B^2} \right) = \frac{d}{dt} \left( \frac{m_i}{2B^2}\frac{\boldsymbol{\omega}}{n}\cdot \boldsymbol{\omega}\right) + \frac{1}{n}\frac{m_i} {2B^2}\frac{\boldsymbol{\omega}}{n}\cdot \boldsymbol{\omega} \frac{d n}{dt}= \\
    =\frac{d \mathcal{K}_\perp}{dt} + \frac{\mathcal{K}_\perp}{n}\frac{dn}{dt} \hspace{9.4em}.
\end{aligned}
\end{equation}

Now, using the leading order definition of the advective derivative for the ions ($d/dt_i = \partial/\partial t + (\boldsymbol{V}_E + \boldsymbol{V}_{di} + \boldsymbol{V}_{\parallel i})\cdot \nabla $), we introduce the continuity equation (\ref{eqn:continuityelectronfinal}) in order to remove the derivative of density and retrieve the term for the final conservation law

\begin{equation}
\begin{aligned}
    \frac{d \mathcal{K}_\perp}{dt} + \frac{\mathcal{K}_\perp}{n}\frac{dn}{dt} = \frac{d \mathcal{K}_\perp}{dt} - \mathcal{K}_\perp \nabla \cdot (\boldsymbol{V}_E+\boldsymbol{V}_{di}+\boldsymbol{V}_{\parallel i})\\
    +\frac{\mathcal{K}_\perp}{n}\left[-\nabla \cdot (n \boldsymbol{V}_{pol})-\frac{D_\Omega}{B^2} \nabla_\perp^2 \Omega +S_n + S_n^ {in} +D_n \nabla_\perp^2 n)  \right]\\ = \frac{\partial \mathcal{K}_\perp}{\partial t} + \nabla \cdot [(\boldsymbol{V}_E+\boldsymbol{V}_{di}+\boldsymbol{V}_{\parallel i})\mathcal{K}_\perp ] - 2 \mathcal{K}_\perp \nabla \cdot (\boldsymbol{V}_E+\boldsymbol{V}_{di}+\boldsymbol{V}_{\parallel i})\\
    +\frac{\mathcal{K}_\perp}{n}\left[-\nabla \cdot (n \boldsymbol{V}_{pol})-\frac{D_\Omega}{B^2} \nabla_\perp^2 \Omega +S_n + S_n^ {in} +D_n \nabla_\perp^2 n)  \right].
    \label{eqn:partialresult2}
\end{aligned}
\end{equation}

Inserting equation \eqref{eqn:partialresult2} into (\ref{eqn:appendparresult3}), we finally find:

\begin{equation}
\begin{aligned}
    \boldsymbol{J}_{pol}\cdot\frac{\boldsymbol{\omega}}{n} =-\frac{\partial \mathcal{K}_\perp}{\partial t} - \nabla \cdot [(\boldsymbol{V}_E+\boldsymbol{V}_{di}+\boldsymbol{V}_{\parallel i})\mathcal{K}_\perp ]\\
    -\frac{\mathcal{K}_\perp}{n}\left[-\nabla \cdot (n \boldsymbol{V}_{pol})-\frac{D_\Omega}{B^2} \nabla_\perp^2 \Omega +S_n + S_n^ {in} +D_n \nabla_\perp^2 n)  \right]
    \\+ 2 \mathcal{K}_\perp \nabla \cdot (\boldsymbol{V}_E+\boldsymbol{V}_{di}+\boldsymbol{V}_{\parallel i})
    +e \boldsymbol{V}_{\boldsymbol{\pi}_i}\cdot \boldsymbol{\omega}
    \label{eqn:appendparresult4}
\end{aligned}
\end{equation}

\noindent
which, after noting that $e \boldsymbol{V}_{\boldsymbol{\pi}_i}\cdot \boldsymbol{\omega}=-(\boldsymbol{V}_E+\boldsymbol{V}_{di})\cdot (\nabla \cdot \boldsymbol{\pi}_i)$, the result used in equation (\ref{partialresult4}) is achieved:

\begin{equation}
\begin{aligned}
    \boldsymbol{J}_{pol}\cdot\frac{\boldsymbol{\omega}}{n} =-\frac{\partial \mathcal{K}_\perp}{\partial t} - \nabla \cdot [(\boldsymbol{V}_E+\boldsymbol{V}_{di}+\boldsymbol{V}_{\parallel i})\mathcal{K}_\perp ]\\
    -\frac{\mathcal{K}_\perp}{n}\left[-\nabla \cdot (n \boldsymbol{V}_{pol})-\frac{D_\Omega}{B^2} \nabla_\perp^2 \Omega +S_n + S_n^ {in} +D_n \nabla_\perp^2 n)  \right]
    \\+ 2 \mathcal{K}_\perp \nabla \cdot (\boldsymbol{V}_E+\boldsymbol{V}_{di}+\boldsymbol{V}_{\parallel i})
    -(\boldsymbol{V}_E+\boldsymbol{V}_{di})\cdot (\nabla \cdot \boldsymbol{\pi}_i).
    \label{eqn:appendparresult5}
\end{aligned}
\end{equation}

\section{GBS normalized units} \label{AppendixA}

In GBS \citep{Giacomin22}, the density and temperatures are normalized to the reference upstream separatrix values $n_0$, $T_{e0}$ and $T_{i0}$ while the electrostatic potential is normalized to $\phi_0=T_{e0}/e$. Velocities are normalized to the reference sound speed $c_{s0}=\sqrt{T_{e0}/m_i}$ and the magnetic field is normalized to the modulus $B_0$ at the magnetic axis. Parallel lengths are normalized to the device major radius $R_0$ and perpendicular lengths to the ion sound Larmor radius $\rho_{s0}=c_{s0}/\Omega_{ci}$. Numerical diffusion coefficients are normalized to $D_0 =  c_{s0} \rho_{s0}^2/R_0$. Finally, time is normalized to $t_0=R_0/c_{s0}$. The reference parameters $n_0$, $T_{e0}$, $T_{i0}$, $R_0$, $B_0$ and $m_i$ set the dimensionless parameters regulating the system, namely the normalized ion sound Larmor radius $\rho_*=\rho_{s0}/R_0$, the ion-electron temperature ratio $\tau=T_{i0}/T_{e0}$, the normalized Spitzer electrical conductivity $\sigma=\sigma_0 T_e^{3/2}$ \citep{Braginskii65} with

\begin{equation}
    \sigma_0 = \frac{5.88}{4 \sqrt{2 \pi}} \sqrt{\frac{m_i}{m_e}}\frac{(4\pi \varepsilon_0)^2}{Z^2 e^4}\frac{T_{e0}^{2}}{\lambda n_0 R_0},
\end{equation}

\noindent
where $\lambda$ is the Coulomb logarithm, the normalized electron and ion parallel heat conductivities $\chi_{\parallel}^e = \chi_{\parallel 0}^e T_{e}^{5/2}$ and $\chi_{\parallel}^i = \chi_{\parallel 0}^i T_{i}^{5/2}$ \citep{Braginskii65} with

\begin{equation}
     \chi_{\parallel 0}^e = \frac{2.37}{\sqrt{2\pi}} \sqrt{\frac{m_i}{m_e}} \frac{(4\pi \varepsilon_0)^2}{Z^2 e^4}\frac{T_{e0}^2}{\lambda n_0 R_0}
\end{equation}

\noindent
and

\begin{equation}
     \chi_{\parallel 0}^i = \frac{2.925}{\sqrt{2\pi}}\frac{(4\pi \varepsilon_0)^2}{Z^4 e^4}\frac{T_{e0}^2 \tau^{5/2}}{\lambda n_0 R_0},
\end{equation}

\noindent
and the normalized ion and electron parallel viscosities $\eta_{\parallel}^e = \eta_{\parallel 0}^e T_{e}^{5/2}$ and $\eta_{\parallel}^i = \eta_{\parallel 0}^i T_{i}^{5/2}$ \citep{Braginskii65} with

\begin{equation}
     \eta_{\parallel 0}^e = \frac{0.5475}{\sqrt{2\pi}} \sqrt{\frac{m_e}{m_i}} \frac{(4\pi \varepsilon_0)^2}{Z^2 e^4}\frac{T_{e0}^2}{\lambda n_0 R_0}
\end{equation}

\noindent
and

\begin{equation}
     \eta_{\parallel 0}^i = \frac{0.72}{\sqrt{2\pi}}\frac{(4\pi \varepsilon_0)^2}{Z^4 e^4}\frac{T_{e0}^2 \tau^{5/2}}{\lambda n_0 R_0}.
\end{equation}

\bibliographystyle{jpp}

\bibliography{jpp-instructions}

\end{document}